**A multiscale model of vascular function in chronic thromboembolic pulmonary hypertension**

Mitchel J. Colebank[1], M. Umar Qureshi[1], Sudarshan Rajagopal[2], Richard Krasuski[3], Mette S. Olufsen[1*]

[1]*Department of Mathematics, North Carolina State University, Raleigh, NC 27695, USA*

[2]*Division of Cardiology, Department of Medicine, Duke University Medical Center, Durham, NC 27710*

[3]*Department of Cardiovascular Medicine, Duke University Health System, Durham, NC 27710*

Running title: A multiscale model of CTEPH

[*]**Corresponding author**

Mette S. Olufsen

Department of Mathematics

North Carolina State University

2311 Stinson Dr.

Raleigh, NC 27607

Email: msolufse@ncsu.edu

**ABSTRACT**

Chronic thromboembolic pulmonary hypertension (CTEPH) is caused by recurrent or unresolved pulmonary thromboemboli, leading to perfusion defects and increased arterial wave reflections. CTEPH treatment aims to reduce pulmonary arterial pressure and reestablish adequate lung perfusion, yet patients with distal lesions are inoperable by standard surgical intervention. Instead, these patients undergo balloon pulmonary angioplasty (BPA), a multi-session, minimally invasive surgery that disrupts the thromboembolic material within the vessel lumen using a catheter balloon. However, there still lacks an integrative, holistic tool for identifying optimal target lesions for treatment. To address this insufficiency, we simulate CTEPH hemodynamics and BPA therapy using a multiscale fluid dynamics model. The large pulmonary arterial geometry is derived from a computed tomography (CT) image, whereas a fractal tree represents the small vessels. We model ring- and web-like lesions, common in CTEPH, and simulate normotensive conditions and four CTEPH disease scenarios; the latter includes both large artery lesions and vascular remodeling. BPA therapy is simulated by simultaneously reducing lesion severity in three locations. Our predictions mimic severe CTEPH, manifested by an increase in mean proximal pulmonary arterial pressure above 20 mmHg and prominent wave reflections. Both flow and pressure decrease in vessels distal to the lesions and increase in unobstructed vascular regions. We use the main pulmonary artery (MPA) pressure, a wave reflection index, and a measure of flow heterogeneity to select optimal target lesions for BPA. In summary, this study provides a multiscale, image-to-hemodynamics pipeline for BPA therapy planning for inoperable CTEPH patients.

**NEW & NOTEWORTHY**

*This paper presents novel computational framework for predicting pulmonary hemodynamics in chronic thromboembolic pulmonary hypertension. The mathematical model is used to identify the optimal target lesions for balloon pulmonary angioplasty, combining simulated pulmonary artery pressure, wave intensity analysis, and a new quantitative metric of flow heterogeneity.*

**INTRODUCTION**

Pulmonary hypertension (PH), defined as a mean pulmonary arterial pressure (mPAP) > 20 mmHg at rest (54), degrades the pulmonary vasculature and, if untreated, progressively leads to right ventricular dysfunction (65). PH encompasses five etiologies, yet only chronic thromboembolic pulmonary hypertension (CTEPH, group 4) is curable via surgical intervention (25, 38). A less invasive approach, known as balloon pulmonary angioplasty (BPA), is utilized if lesions are too distal. CTEPH affects 3--30 people per million worldwide (33), but its prevalence is underestimated due to the common symptoms of dyspnea and fatigue (40). CTEPH, typically preceded by an acute pulmonary embolism, involves recurrent obstruction of the pulmonary arteries, leading to perfusion deficits and pulmonary vascular remodeling. Besides elevated mPAP, clinical manifestations of CTEPH include increased flow heterogeneity and elevated right ventricular afterload (17, 39).

Disease diagnosis requires both static imaging, such as computed tomography (CT) imaging and ventilation-perfusion (V/Q) scans, as well as dynamic pressure readings via right heart catheterization. CTEPH diagnosis requires at least one large uni- or bilateral perfusion defect, identified on a V/Q scan or CT image (38, 67). Recent studies using optical coherence tomography (22, 25) sub-classified CTEPH lesions as (1) ring-like, (2) web-like, (3) subtotal occlusions, (4) total occlusions, and (5) tortuous lesions. Of these, subtotal occlusions, ring-like, and web-like lesions are the most common and have the lowest probability of complication in response to BPA intervention (25).

Current hypotheses suggest that flow redistribution disrupts homeostatic vascular stretch and shear stress in both obstructed and unobstructed vascular pathways (32), causing secondary pulmonary arteriopathy (14, 27, 31). Muscularization and narrowing of small vessels raise PVR, causing pronounced wave reflections, increased mPAP (53, 63), and eventual large artery dilatation (51, 66). These features are typical in PAH as well as in CTEPH, and both PAH and CTEPH patients have similar PVR and pulmonary vascular compliance values (15). However, PAH initiates with distal vessel remodeling, whereas CTEPH is onset by large artery emboli, causing distal vessel remodeling in later stages of the disease (15).

Pulmonary thromboendarterectomy (PTE) is the gold standard therapeutic intervention for CTEPH with a five-year survival rate > 83% (47), yet 12% to 60% of patients are inoperable due to distal lesions in the segmental and subsegmental arteries (24, 25). The alternative treatment for these patients is BPA therapy, a less invasive procedure that mechanically disrupts thrombi using a balloon catheter (18). Complete BPA therapy requires multiple interventions over several sessions (31, 67). Variability in lesion morphology, severity, and location influence BPA outcome (36) making lesion selection nontrivial for physicians and motivating development of a patient-specific planning tool. Though fractional flow reserve (FFR) can quantify a lesion's impact on hemodynamics (25), most clinics only utilize mPAP and V/Q scans to determine whether BPA is successful, necessitating additional BPA planning indices. CTEPH incidence is expected to grow in the coming decade, hence predictive tools integrating multimodal data will be crucial to improve the safety, efficacy, and efficiency of BPA.

Large pulmonary arteries can be analyzed and reconstructed from CT, yet limited image resolution makes it impossible to extract patient-specific small arterial geometries. However, the structure and function of small pulmonary arterial vessels are imperative in understanding CTEPH and its treatment. Clinical studies demonstrate a negative correlation between long-term survival rates and elevated PVR post-PTE (19). In addition, Jujo et al. (23) conclude that severe arteriopathy in pulmonary arterioles is linked to recurrent CTEPH after PTE. Imaging limitations encourage developing new tools, such as computational modeling, for assessing small vessel function and its link to proximal artery hemodynamics.

Numerous studies use image-based CFD modeling to predict hemodynamics in large arteries (56, 69, 71). Spazzapan et al. (56) integrate CT pulmonary angiography images with a three-dimensional (3D) CFD model, showing that low magnitude wall shear stress (WSS) distal to LPA and RPA stenoses increases in post-PTE simulations. The patient-specific 3D CFD study by Yang et al. (69) also predicts proximal pulmonary arterial WSS, describing a negative correlation between PH severity and proximal artery WSS magnitude in pediatric patients. Several studies use less computationally intensive 1D CFD models to predict wave propagation and wave reflections in the large vessels. Yin et al. (71) predict energy losses due to stenoses in the coronary arteries, finding a maximum error of 3% between the 1D and 3D CFD FFR predictions. Likewise, Bezerra et al. (4) predict nearly identical FFR from 3D and 1D models for multiple coronary artery patient geometries. Most studies lump the entire distal vasculature as a boundary condition, e.g., using a Windkessel boundary condition (56, 71) that represents the microvasculature as an RCR circuit. While this is adequate for examining large vessel dynamics, modeling distal arteriopathy requires a multiscale approach including the large and small vessels. Several studies address this by solving simplified CFD equations in a morphological tree of the distal vessels (11, 57). Spilker et al. (57) couple a Strahler based (20) distal vascular bed model to a 1D CFD model to predict and match pulmonary arterial hemodynamics to porcine and human data. Clipp and Steele (11) use the structured-tree boundary condition, originally developed by Olufsen (42), with their 1D CFD model to predict pulmonary arterial hemodynamics during inspiration and expiration in lambs.

Several studies predict both large and small vessel hemodynamics (8, 10, 43, 69). Yang et al. (69) couple the distal vasculature model from Spilker et al. (57) to a 3D pulmonary arterial CFD model, showing increased arteriolar WSS in PAH versus normotension. Clark and Tawhai (10) predict pulmonary arterial and venous hemodynamics using a wave-transmission model coupled to a capillary sheet model. Their multiscale study provides qualitative predictions of lung perfusion, with reduced pulmonary vascular compliance corresponding to increased flow heterogeneity. The study by Olufsen et al. (43) solves inertially driven 1D equations in the large pulmonary arteries and a linearized wave-equation in the arterioles of the structured tree model, illustrating that pulmonary microvascular rarefaction and increases mPAP. Similarly,

Chambers et al. (8) predict hemodynamics in normotensive and PH mice, using micro-CT imaging for the large arteries, whereas the small arterial networks are represented by structured trees informed by high-resolution imaging data.

To our knowledge, no previous studies have modeled CTEPH by incorporating pressure-loss models nor do they attempt to predict large and small vessel fluid dynamics in CTEPH. This study addresses this critical gap in the literature using a novel multiscale approach. We simulate large artery dynamics in a patient-specific geometry extracted from CT images, while small vessel fluid dynamics are predicted in asymmetric, structured trees. We combine this computational model with two energy loss models for ring- and web-like thromboembolic lesions to predict pressure, flow, and area in normotensive and CTEPH scenarios. To understand lesion and vascular remodeling effects on hemodynamics, we compute mPAP, FFR, a wave reflection index, and lung perfusion. The latter is converted to a quantitative flow heterogeneity index utilizing the Kullback-Leibler (KL) divergence (3). Finally, we simulate "virtual BPA" by modulating lesion severity and identify the most significant set of lesions for percutaneous intervention. This study develops a proof-of-concept CTEPH model, which can assist teams of cardiologists, pulmonologists, and radiologists in determining optimal intervention strategies for BPA therapy. Similar to other computational tools (5, 28, 69), our model can potentially assist in the planning and treatment of CTEPH and other vascular occlusive disorders.

**METHODS**

*Network Geometry*

In this study, the pulmonary arterial geometry consists of an image-based large artery network generated from a CT image and a self-similar, fractal-tree representing the small arteries and arterioles.

*Large vessel geometry.*

We generate a network of large pulmonary arteries from a publicly available chest CT scan (from the vascular model repository (68)[1]). The CT image is from a healthy, 67-year-old female volunteer, captured using Omnipaque 350 contrast agent. Similar to our previous study (8), we segment and reconstruct the pulmonary arterial network and lung cavity using the open-source segmentation software 3DSlicer[2] (16). When segmenting the geometry, we first identify the main (MPA), left (LPA), and right (RPA) pulmonary artery, and then manually segment all lobar, segmental, and subsegmental vessels. We generate the 1D network by reducing the 3D vessel geometry to centerlines using the Vascular Modeling ToolKit (VMTK) (2) and use our network extraction algorithm to generate a directed graph from the centerlines (12). This algorithm converts the network to a series of arcs (vessels) and nodes (junctions), encoded by a connectivity matrix. For non-terminal arteries, each vessel's outlet is connected to either two or three vessel inlets, deemed the parent and daughter vessels, respectively. Terminal vessels are prescribed a structured tree model at their outlet, as defined below. Each vessel has a constant radius and a length. Radii are computed as the mean within the center 80% of the vessel, and the lengths are computed as the Euclidean distance between each point in the vessel. Figure 1 illustrates this process showing three planes in the CT image, the segmented network and lung cavity, and the 1D network.

---

[1] http://simvascular.github.io/

[2] http://www.slicer.org

**FIGURE 1**

**Figure 1:** Schematic of model development, starting from image analysis and image segmentation and ending in construction of the skeletonized network and computational domain.

*Small vessel geometry.*

It is difficult to obtain high-fidelity measurements of the small arteries and arterioles from CT images. To circumvent this, we represent the small vessels with fractal, self-similar structured trees (42, 43). We assume that each daughter vessel is related to its parent via the radius scaling factors $\alpha$ and $\beta$, and a length to radius ratio $\ell_{rr}$. The dimensions of any vessel in the structured tree can be expressed in terms of the terminal large artery radius $r_{term}$ (cm) proximal to the beginning of the structured tree

$$r_{n,m} = r_{term}\alpha^n\beta^m, \ \ L_{n,m} = \ell_{rr}\, r_{n,m}. \tag{1}$$

Following (50), who use a combination of pulmonary artery literature data, the length to radius ratio used in this study is

$$\ell_{rr} = \begin{cases} 15.75\, r^{1.10}, & r \geq 0.005 \\ 1.79\, r^{0.47}, & r < 0.005. \end{cases} \tag{2}$$

The structured tree is terminated after a specified minimal radius, $r_{min}$, has been reached. We use parameter values from a previous investigation from high-resolution micro-CT scans of control and hypertensive mice (8), and set $\alpha = 0.88$ and $\beta = 0.68$. The terminal radius of the structured tree is based on anatomical literature and set to 7.5$\mu$m (61).

*Fluid dynamics*

The 1D model consists of the large pulmonary arteries segmented from the CT image, and the small pulmonary arteries and arterioles represented by structured trees. In the large arteries, we predict inertia-driven hemodynamics using a system of nonlinear partial differential equations (PDEs), while fluid dynamics in the network of small vessels are predicted by solving a linearized wave equation.

*Large vessel hemodynamics.*

Similar to previous studies (13, 48, 50), we compute flow $q(x,t)$ (ml/s), pressure $p(x,t)$ (mmHg), and cross-sectional area $A(x,t)$ (cm²) in each large vessel by enforcing conservation of volume

$$\frac{\partial A}{\partial t} + \frac{\partial q}{\partial x} = 0, \tag{3}$$

and the balance of momentum

$$\frac{\partial q}{\partial t} + \frac{\partial}{\partial x}\left(\frac{q^2}{A}\right) + \frac{A}{\rho}\frac{\partial p}{\partial x} = -\frac{2\pi\nu R}{\delta}\frac{q}{A}, \tag{4}$$

where $\rho = 1.055$ (g/mL) is the blood density and $\nu$ (cm²/s) is the kinematic viscosity. The right-hand side of equation (4) is derived under the assumption that the fluid velocity profile, $u_x(r,x,t) = q/A$ (cm/s), has a linearly decreasing boundary layer thickness $\delta$ (cm), i.e.,

$$u_x = \begin{cases} \bar{u}_x, & 0 \leq r \leq R - \delta \\ \bar{u}_x \dfrac{R-r}{\delta}, & R - \delta \leq r \leq R, \end{cases} \tag{5}$$

where $\bar{u}_x = q/A$. The boundary layer $\delta$ is approximated by $\sqrt{\nu T/2\pi}$, where $T$ (s) is the cardiac cycle length (35). Under this assumption, the shear stress in the large vessels is

$$\tau(x,t) = -\mu_{la}\left[\frac{\partial u_x}{\partial r}\right]_{r=R} = \mu_{la}\frac{\bar{u}_x}{\delta}, \tag{6}$$

where $\mu_{la} = 0.032$ (g/cm/s) is the large artery blood viscosity. To close the system of equations, we impose a pressure-area relation governing the interaction between the vascular wall and the blood. Under the assumption that vessels are thin-walled, linearly elastic, homogenous, and isotropic (42, 71), we get

$$p(x,t) = p_0 + \frac{4}{3}\frac{Eh}{r_0}\left(\sqrt{\frac{A}{A_0}} - 1\right), \tag{7}$$

where $p_0$ (mmHg) is the reference pressure at which $A = A_0 = \pi r_0^2$, $E$ (mmHg) is Young's modulus in the circumferential direction, and $h$ (cm) is the wall thickness. For clarity, simulations are conducted in cgs units, but the results are provided in mmHg using the conversion 1 mmHg = 1333.22 g/cm/s$^2$.

Previous investigations (30) suggest that pulmonary vascular wall properties, e.g., Young's modulus ($E$) and wall thickness ($h$), change with vessel caliber. To account for these varying wall properties, we assume that the wall-stiffness $Eh/r_0$ can be represented by

$$\frac{Eh}{r_0} = k_1^{la} e^{r_0 k_2^{la}} + k_3^{la}, \tag{8}$$

as described by Olufsen (42). Here $k_1^{la}$ (mmHg), $k_2^{la}$ (cm$^{-1}$), and $k_3^{la}$ (mmHg) are stiffness parameters for the large arteries. Normotensive values of $k_1^{la} = 2.5 \times 10^6$, $k_2^{la} = -15$, and $k_3^{la} = 6.4 \times 10^4$ are based on our previous investigation (48, 50).

The nonlinear system of PDEs is hyperbolic with characteristics pointing in the opposite direction; thus, we require boundary conditions at each vessel inlet and outlet ($x = 0$ and $x = L$, respectively). We digitize an average control flow waveform, available from Simvascular, using GraphClick[3] to provide an inflow boundary condition at the inlet of the MPA. The cardiac output is scaled to match the flow values used in (43). At non-stenotic vessel junctions, we prescribe continuity of static pressure and conservation of flow

$$p_p(L_p, t) = p_{d_1}(0, t) = p_{d_2}(0, t), \tag{9}$$

$$q_p(L_p, t) = q_{d_1}(0, t) + q_{d_2}(0, t), \tag{10}$$

which holds $\forall t \in [0, T]$. Lastly, we couple the terminal large arteries to the structured tree model, as described later.

*Stenosis models.*

As discussed earlier, CTEPH progression is attributed to multiple emboli and vascular proliferation, which are grouped into five types (25): ring-like, web-like, subtotal occlusion, total occlusive, and tortuous lesions.

---

[3] http://www.arizona-software.ch/graphclick/

In this work, we model ring-like lesions (accounting for approximately 30% of all lesions) and web-like lesions (accounting for approximately 64% of all lesions) (25).

We model the former lesion type using the pressure loss term derived from Young and Tsai's work (72)

$$\Delta p_{ring} = \frac{\mu_{la} K_v}{2\pi r_p^3} q + \frac{\rho K_t}{2 A_p^2} \left( \frac{A_p}{A_s} - 1 \right)^2 |q| q + \frac{\rho K_u L_s}{A_p} \frac{\partial q}{\partial t}, \tag{11}$$

where $K_v, K_t,$ and $K_u$ (dimensionless) are the loss coefficients due to viscous, turbulent, and inertial forces, respectively. Moreover, $r_p$ (cm) and $A_p$ (cm$^2$) denote the unobstructed radius and area proximal to the lesion, and $L_s$ is the length of the ring-like lesion. This stenosis model has been implemented in 1D models and validated against 3D models in the systemic circulation (4, 45, 71). The viscous loss $K_v$, related to the stenosis geometry, is given by (71)

$$K_v = 16 \frac{L_s}{r_p} \left( \frac{A_p}{A_s} \right)^2. \tag{12}$$

These lesions are typically located proximal to junctions (25) and are added to equation (9) to account for energy losses in the daughter branches.

We model web-like occlusions as semi-porous structures, which only partially occlude the vascular lumen (21). The pressure drop in web-like lesions is modeled as

$$\Delta p_{web} = L_w \left( \frac{\mu_{la}}{K_{perm}} \frac{q}{A} + \rho K_w \left( \frac{q}{A} \right)^2 \right), \tag{13}$$

where $L_w$ (cm) is the length of the obstructed area, $K_{perm}$ (cm$^2$) is the permeability of the web-like lesion, and $K_w$ (dimensionless) is the pressure drop due to inertial effects. Equation (13), the Darcy-Forchheimer equation, is used for fluid flow through porous media at moderate Reynold's numbers (i.e. > 10) (73). Previous work (29, 73) assume that $K_w \approx G / K_{perm}^{1/2}$, where $G$ (cm) depends on the material in question. Since material properties are unknown, we assume $G = 0.10$ for all simulations involving web-like lesion and place them at the vessel midpoint as these lesions can extend throughout a single vessel (21). Since web-like lesions can occupy a majority of a vessel (25), we place them at the midpoint of the vessel, measuring 50% of the vessel length $L$, i.e. for $0.25L \leq x \leq 0.75L$.

We quantify pressure-losses due to lesions using the FFR. Similar to coronary studies (5), the FFR is calculated as

$$\text{FFR} = \frac{\bar{p}_{dist}}{\bar{p}_{prox}}, \tag{14}$$

where $\bar{p}_{prox}$ and $\bar{p}_{dist}$ are the mean pressures measured immediately proximal and distal to the lesion.

*Small vessel hemodynamics.*

While inertial effects dominate large vessel hemodynamics, blood flow in the small arteries and arterioles is dominated by viscous forces. For this reason, we solve linearized forms of equations (3) and (4). The linearized equations are converted to a discrete, frequency domain Fourier series, resulting in a wave-equation. This is derived in the Appendix, and leads to analytical pressure and flow solutions, $P(x, \omega)$ and $Q(x, \omega)$, for each frequency $\omega$, and are related via the impedance, $Z(x, \omega) = P(x, \omega)/Q(x, \omega)$. The input impedance at $x = 0$ can then be written as

$$Z(0, \omega) = \frac{i g_\omega^{-1} \sin(\omega L/c) + Z(L, \omega) \cos(\omega L/c)}{\cos(\omega L/c) + i g_\omega Z(L, \omega) \sin(\omega L/c)}, \tag{15}$$

for any frequency $\omega \neq 0$, whereas the zero-frequency value, $Z(0,0)$, analogous to the DC component in electrical circuit theory, is given by

$$\lim_{\omega \to 0} Z(0, \omega) = \frac{8\mu \ell_{rr}}{\pi r_0^3} + Z(L, 0). \tag{16}$$

The above equations depend on the frequency domain quantities $g_\omega$ and $c$, which are defined and relegated to the Appendix. These terms depend on $Eh/r_0$, which has the same form as equation (8), but with parameters $k_1^{sa}$, $k_2^{sa}$, and $k_3^{sa}$.

Noting that $Q(L, \omega) = P(L, \omega)/Z(L, \omega)$ for each frequency, we can predict impedance, pressure, and flow using

$$P(L, \omega) = P(0, \omega) \cos(\omega L/c) - Q(0, \omega) \frac{i}{g_\omega} \sin(\omega L/c), \tag{17}$$

$$Z(L, \omega) = \frac{i g_\omega^{-1} \sin(\omega L/c) - Z(0, \omega) \cos(\omega L/c)}{g_\omega \cos(\omega L/c) Z(0, \omega) - \cos(\omega L/c)}, \tag{18}$$

for $\omega \neq 0$. For the zeroth frequency, the solution is

$$P(L, 0) = P(0,0) - Q(0,0) \frac{8\mu \ell_{rr}}{r_0^3}, \tag{19}$$

$$Z(L,0) = Z(0,0) - \frac{8\mu\ell_{rr}}{r_0^3}. \tag{20}$$

This enables us to predict pressure and flow in any daughter vessel provided the pressure and flow of the parent are known (43).

The non-Newtonian effects of blood viscosity $\mu$ (and hence kinematic viscosity $\nu$) become more influential as the vessel radii decrease towards the arterioles. Following (46) and our previous study (8), we model the viscosity in the small arteries and arterioles by a nonlinear radius dependent function of the form

$$\mu^*(r_0) = \mu_{la}[1 + (\mu_{0.45} - 1)\mathcal{D}]\mathcal{D}, \tag{21}$$

$$\mu_{0.45}(r_0) = 6e^{-0.17r_0} + 3.2 - 2.44e^{-0.12r_0^{0.645}}, \tag{22}$$

where $\mu_{la}$ is the large artery viscosity, $\mathcal{D} = [2r_0/(2r_0 - 1.1)]^2$ and $\mu_{0.45}(r_0)$ is the relative viscosity at an average hematocrit level of 0.45 (46). We assume a typical healthy blood hematocrit of 0.45.

We assume continuity of pressure and conservation of flow at each junction in the small vessels of the structured tree. Under these assumptions, the impedance at a junction parallels resistors in parallel, i.e., $Z_p^{-1} = Z_{d_1}^{-1} + Z_{d_2}^{-1}$. At the terminating end of the structured tree (i.e., $Z(L,0)$ at the last generation branches), the terminal impedance $Z_{trm}$ = constant is prescribed. Since the structured tree results in thousands of vessels, the flow at the end of the structured trees are on the order of $10^{-5} - 10^{-8}$. For this reason, the capillary pressure, $P_{trm} = Z_{trm} Q_{trm}$, requires $Z_{trm}$ to be on the order of $10^5 - 10^9$ to achieve non-negligible pressure values. Once we reach the terminal structured tree branch, all proximal impedance values in the tree are computed. Knowing the impedance at the root of the tree provides an interface between the large and small vessels. The large artery predictions are then used to simulate structured tree hemodynamics using the impedance relations above.

*Numerical Solution.*

Large vessel hemodynamics (equations (3), (4), and (7)) are solved using the Ritchmeyer two-step Lax-Wendroff scheme (42), which is second order accurate in space and time. At non-stenotic junctions, pressure continuity and flow conservation are enforced via a Newton-Raphson root finding routine. Pressure losses due to ring-like lesions, located proximal to bifurcations, are also solved using a Newton-Raphson routine. For web-like lesions, located at vessels' midpoint, we minimize residual equations using gradient-descent with a backtracking line search algorithm (26). To increase numerical accuracy at the junctions and lesion locations, we derive and utilize the analytical Jacobian of the residuals.

To couple the large vessels to the structured tree, the frequency impedance is convolved with the pressure at the end of the large arteries. This is numerically approximated as

$$q_L^n = \Delta t \sum_{k=1}^{n} p_L^{n-k} \left(Z^k\right)^{-1},$$ (23)

where $q_L^n$ and $p_L^{n-k}$ are the numerical solutions of flow and pressure at the end of the large terminal arteries, and $Z^k$ is the vascular impedance of the whole structured tree calculated using equation (15).

*Wave intensity analysis.*

Wave intensity analysis (WIA) is a time-domain decomposition of forward and backward waves and is utilized in understanding the progression of cardiovascular disease (48, 50, 59). The so-called pressure and velocity "wavefronts" (44) $\delta p^\pm$ and $\delta u^\pm$ are calculated as

$$\delta p^\pm = \frac{\delta p \pm \rho c(p)\,\delta u}{2}, \qquad \delta u^\pm = \frac{1}{2}\left(\delta u \pm \frac{\delta p}{\rho c(p)}\right),$$ (24)

where the notation $\pm$ indicates the forward ("+") or backward ("-") running wave components, and $\delta p$ and $\delta u$ are the incremental changes in pressure and velocity, respectively. The wave speed $c(p)$ is calculated from the pressure-area relation

$$c(p) = \sqrt{\frac{A}{\rho}\frac{\partial p}{\partial A}} = \sqrt{\frac{Eh}{r_0}\frac{1}{2\rho}}\left(\frac{A}{A_0}\right)^{1/4}.$$ (25)

Both forward and backward waves are classified as compression ($\delta p^\pm > 0$) or decompression $\delta p^\pm < 0$) waves. Forward compression waves (FCW) originate from the right ventricle, increasing pressure and accelerating flow as blood travels towards the periphery. Forward expansion waves (FEW) also originate in the proximal arteries but decrease the pressure and flow velocity. In contrast, backward compression waves (BCW) and expansion waves (BEW) propagate from the periphery, the former increasing pressure while decelerating flow, while the latter does the opposite. The wave intensity $W^{I\pm}$ is calculated as a product of the velocity and pressure wavefronts. Since wavefronts are not continuous but discrete approximations, we use the time-normalized wave intensity

$$W^{I_\pm} = \frac{\delta p^\pm}{\delta t}\frac{\delta u^\pm}{\delta t}.$$ (26)

After determining the type and nature of the waves, a wave reflection coefficient, $\mathcal{R}$, can be calculated. In this study, we use

$$\mathcal{R} = \frac{\Delta p_+}{\Delta p_-}, \tag{27}$$

where $\Delta p_+$ and $\Delta p_-$ are the sums of the backward and forward compression waves, respectively (41).

*Wave intensity analysis.*

To predict perfusion throughout the lung, we map 1D flow predictions to the lung tissue. First, we segment the lung tissue using thresholding and cutting tools in 3D Slicer. This segmentation is converted to a volumetric mesh, and exported to MATLAB (Mathworks, Natick, MA). Using the 3D coordinates of the vascular centerlines, we map sections of the lung tissue to the distal ends of terminal vessels using the nearest-neighbor algorithm in MATLAB. Flow predictions in all terminal vessels are mapped to the corresponding nearest tissue, providing 3D lung perfusion. These predictions are converted to a probability distribution function (PDF) using MATLAB's kernel density estimation function, ***ksdensity***. To quantify flow heterogeneity in control, CTEPH, and post-surgical intervention simulations, we compute the KL divergence, a measure of relative entropy between two probability distributions (3), for the flow PDFs.

The KL divergence is a probabilistic measure between two distributions (3). Let $\mathcal{A}$ and $\mathcal{B}$ be discrete probability distributions defined on some probability space $\mathcal{W}$, where $\mathcal{A}$ is our reference or target distribution. Then, the KL-divergence is defined as

$$D_{KL}(\mathcal{A}|\mathcal{B}) = \sum_{w \in \mathcal{W}} \mathcal{A}(w) \log\left(\frac{\mathcal{B}(w)}{\mathcal{A}(w)}\right), \tag{28}$$

where $D_{KL} \to 0$ indicates identical distributions, while increasing values of $D_{KL}$ indicates a bigger mismatch between the two distributions.

In summary, we compute the perfusion map and quantify flow heterogeneity as follows.

1. Project the 1D flow predictions to the centerline network in 3D.
2. Determine the terminal vessels closest to the segmented lung tissue via a nearest-neighbor algorithm.
3. Prescribe the flow to the entire portion of tissue.
4. Construct the normotensive and CTEPH flow PDFs, $\mathcal{A}$ and $\mathcal{B}$, via kernel density estimation.
5. Compute $D_{KL}(\mathcal{A}|\mathcal{B})$.

*Chronic thromboembolic pulmonary hypertension (CTEPH).*

Results from Kawakami et al. (25) report an average of 20 lesions in patients with inoperable CTEPH. Thus, we model CTEPH by adding 9 ring-like (concentric stenosis) and 11 web-like (longitudinal obstructions within a vessel) lesions to the proximal pulmonary arteries (25). To simulate small vessel disease due to secondary arteriopathy, we introduce arterial vasoconstriction and progressive remodeling of the distal vasculature (51). Moreover, we simulate capillary and venous remodeling by changing the terminal impedance $Z_{trm}$ (19). We set up a normotensive simulation and four disease cases and examine changes in hemodynamic predictions. These disease cases are summarized in Table 1 and parameter values are given in Table 2. To simulate the effects of BPA, we reduce both stenosis and downstream vasoconstriction severity in three lesions and quantify improvements in hemodynamics after BPA.

*Disease cases.*

*Case a: Lesions only.* This scenario represents CTEPH without pulmonary vascular remodeling, illustrating the sole effects of ring- and web-like lesions. Ring-like lesions induce an area reduction of 90%, a length 25% of the original vessel, and are placed proximal to vessel junctions (25). Web-like lesions have a permeability parameter $K_{perm} = 0.009$, a lesion length corresponding to 50% of the original vessel's length, and are placed in the middle of vessels from $0.25L \leq x \leq 0.75L$, where $L$ is the vessel length (25). Vessels that contain either lesion type are tenfold stiffer to account for wall stiffening due to thromboembolic material.

*Case b: Lesions with local remodeling.* In this scenario, the disease has progressed downstream to the lesions. Remodeling narrows the vascular lumen and moderately thickens the vascular walls, decreasing vessel compliance. Parameters $k_1$, $k_2$, and $k_3$ are increased in all vessels (both large and small) distal to the lesion while reducing the reference vessel radius $r_0$ by 40%. We also increase pulmonary vascular resistance by increasing $Z_{trm}$ in terminal vessels that are narrowed.

*Case c: Lesions with small vessel arteriopathy.* Prolonged exposure to CTEPH is associated with secondary arteriopathy throughout the pulmonary vasculature, affecting vessels with cross-sectional areas $\leq 5$ mm$^2$ (31, 51). In particular, long-term elevation in flow and shear stress in unobstructed vessels leads to endothelial dysfunction and increased PVR (55). We model this by inducing global small vessel remodeling, i.e., reducing the reference radius $r_0$ in all vessels with a cross-sectional area $< 5$ mm$^2$ by 40% and increasing vessel stiffness. We also increase $Z_{trm}$ in the structured trees.

*Case d: Lesions with proximal dilation and global small vessel arteriopathy.* In the most severe cases of CTEPH, small vessel arteriopathy causes large artery dilation and stiffening (25, 31). This scenario includes the global small vessel remodeling described in case c, with dilation and thickening of arteries with cross-sectional areas ≥ 10 mm$^2$ by 10%, as documented previously (51). We further increase stiffness and raise PVR in all terminal vessels.

**TABLE 1**

Table 1: Summary of disease scenarios simulated in this study.

**TABLE 2**

Table 2: Parameter values for simulating the five different disease scenarios described in *Chronic thromboembolic pulmonary hypertension (CTEPH)*. $k_3$ and $Z_{trm}$ values are listed for vessels both in unobstructed and obstructed paths (in parentheses).

**TABLE 3**

Table 3: Parameters describing ring- and web-like lesions and their values.

*Balloon pulmonary angioplasty (BPA) therapy.*

Inoperable CTEPH patients not suitable for PTE can instead receive BPA, a less invasive procedure for treating distal vascular obstructions. The goal of BPA is to reperfuse the lung, decrease PVR, and reduce mPAP. We simulate BPA intervention by simultaneously decreasing lesion severity in three locations at a time. BPA does not completely remove the lesions; hence, we still include them in the model but reduce their severity. Specifically, we reduce the stenosed area $A_s$ and lesion length $L_s$ from 90% to 50% for ring-like lesions and increase $K_{perm}$ (see equation (13)) from 0.009 to 0.05 for web-like lesions. We also return vessel reference radii, $r_0$, to 90% of their original value in the obstructed pathways, mimicking decreased vasoconstriction due to lung reperfusion.

Since multiple interventions must be performed to improve hemodynamic function, the most severe lesions are prioritized for treatment to optimize increases in lung perfusion and decreases in mPAP, PVR, and wave reflections. This study uses simulations to select the "optimal" interventional option by quantifying flow heterogeneity, MPA pressure, wave reflection coefficients, along with post-BPA perfusion maps. We perform all virtual procedures in disease case (d) and compare hemodynamic indices pre- and post-BPA.

**RESULTS**

The large pulmonary arterial network includes 226 vessels including the MPA, LPA, and RPA, and all visible segmental and subsegmental branches. Centerlines constructed by VMTK are processed and implemented in the 1D model. With the exception of one trifurcation, all network junctions are bifurcations. Vessels with a length $\leq 1.25$ mm (3 of the 226 vessels) are elongated to match the spatial resolution of the fluid dynamics solver. The mean length and radius are $1.82 \pm 1.28$ cm and $0.23 \pm 0.22$ cm. Maximum and minimum length were 6.42 cm and 0.13 cm, while maximum and minimum radii values were 1.28 cm and 0.05 cm. The average terminal vessel radius was $0.11 \pm 0.04$.

We simulate normotensive hemodynamics and four disease cases with increasing CTEPH severity (cases a-d). CTEPH is induced by including 20 lesions in the segmental and subsegmental vessels and by increasing the stiffness of the large and small arteries. We compare large and small vessel hemodynamics, network perfusion, and wave intensities for all simulations. Using case (d) as our baseline for BPA, we perform virtual treatments by reducing lesion severity and vasoconstriction due to lesion-induced hypoperfusion in three lesions at a time. We select an optimal procedure based on improvements in flow heterogeneity, MPA pressure, and the wave reflection coefficient.

*Effects of lesion severity.*

To understand how lesions affect pulmonary hemodynamics, we increase the stenosis area of ring-like lesions and reduce the permeability of the web-like lesions. We increase the relative stenosis area, $A_s/A_p$, from 0% to 90%, and decrease $K_{perm}$ from 1.5 to 0.003. Figure 2 shows changes in mean MPA pressure along with FFR estimates at all lesion locations presented as box-and-whisker plots. Overall, increasing ring-like lesion severity raises the mean MPA pressure from 12 to $\approx$18 mmHg. Decreasing the permeability of web-like lesions is less influential on MPA pressure, raising it by approximately 0.5 mmHg. The minimum FFR for ring-like lesions is 0.16 when the area reduction is 90%, while the minimum FFR in web-like lesions is 0.66 for $K_{perm} = 0.003$. The average FFR for ring-like lesions at 90% area reduction is 0.25, whereas the average FFR is only 0.86 for the most web-like lesions when $K_{perm} = 0.003$.

We also map flow predictions from the 1D model to the segmented lung tissue. Figure 3 shows changes in perfusion when ring-like lesions' area reduction rise from 0% to 90%. Flow heterogeneity, quantified by the KL-divergence, increases with lesion severity. Results show larger perfusion deficits in the lower and upper lobes than other regions of the lung. Changing $K_{perm}$ in web-like lesions alone does not alter perfusion predictions substantially (results not shown). The flow PDF's, constructed by kernel density estimation, are shown in Figure 3.

**FIGURE 2**

Figure 2: Predictions of mean MPA pressure (top), and fractional flow reserve (FFR, bottom) when increasing lesion severity in ring-like (left) or web-like (right) lesions. FFR predictions correspond to 9 ring-like lesions and 11 web-like lesions. Overall, ring-like lesions have a greater effect on MPA pressure and FFR than web-like lesions.

**FIGURE 3**

Figure 3: Perfusion maps for increasing ring-like stenosis severity. The top left panel (0%) shows the normotensive perfusion map, while the bottom right perfusion map shows 90% severity. The Kullback–Leibler (KL) divergence, constructed from the flow probability density function (PDF), quantifies the flow heterogeneity. PDFs are constructed using kernel density estimation with a bandwidth of 0.009 and a reflection boundary correction.

*Large vessel hemodynamics*

Normotensive and CTEPH predictions of pressure, flow, and shear stress at the midpoint of the MPA, LPA, and RPA are displayed in Figure 4. Simulations from case (a) (normotensive parameters with lesions only) result in a mean MPA pressure of 22.5 mmHg. The downstream stiffening and narrowing in case (b) have a negligible effect on pressure from case (a). Including global small vessel stiffening in case (c) elevates systolic and diastolic pressures slightly. In contrast, severe CTEPH, case (d), raises the pulse pressure (the difference between systolic and diastolic values), with systolic pressures $\approx$ 45 mmHg and diastolic pressures $\approx$ 5 mmHg.

Flow predictions are oscillatory during diastole in normotension and case (a), while increased vessel stiffness, cases (b)-(d), dampens the oscillations. All simulations have a larger flow distribution to the RPA than the LPA; normotensive predictions have a 43/57 flow split to the LPA/RPA, while in severe CTEPH, case (d), the ratio is 22/78. MPA and LPA Shear stress magnitude is greater in normotension than in any of the CTEPH simulations, while RPA shear stress is largest in case (d).

**FIGURE 4**

Figure 4: Pressure, flow, and shear stress in the main, left, and right pulmonary arteries in each disease scenario. Simulations in case (a) include vascular lesions without large or small vessel remodeling. Simulations in cases (b), (c), and (d) include additional local small vessel remodeling, global small vessel remodeling, and large vessel dilatation and stiffening, respectively.

Figure 5 shows pressure predictions throughout the large pulmonary arteries in normotension, case (a), and case (d) at different times within the cardiac cycle. Yellow vessel segments denote lesion locations. Results show a dramatic rise in pressure proximal to lesions for cases (a) and (d), while pressure drops downstream to the lesions. For case (a), the average FFR is 0.33 (range [0.16, 0.95]) for ring-like lesions and 0.95 (range [0.88, 0.94]) for web-like lesions. In case (d), the average FFR is 0.37 (range [0.19, 0.95]) for ring-like lesions and 0.94 (range [0.81, 0.99]) for web-like lesions. Note that the pressure magnitude in vessels downstream to the lesions in case (d) are similar to normotension. Finally, we observe that peak systolic pressure trails maximal MPA flow in all three simulations.

Figure 6 shows pressure and flow predictions along two pathways: one obstructed by lesions and one unobstructed. Pressure and flow results are reduced significantly in vessels distal to the lesion in the obstructed pathway. For case (d), stiffening and narrowing of the small vessels causes a pressure increase downstream to the lesion, yet the pressure has a smaller magnitude compared to normotensive simulations. Besides, the flow is significantly reduced in all CTEPH cases once lesions are included, though the flow does increase slightly in the terminal vessel when moving to case (d).

In contrast, pressure in the unobstructed pathway increases with CTEPH severity. The flow to the terminal vessel increases when lesions are added in case (a), but decreases in cases (c) and (d). Terminal vessel systolic pressure in case (d) is $\approx 45$ mmHg in vessels in the unobstructed path, whereas the terminal systolic pressure is 10 mmHg in the obstructed pathway.

**FIGURE 5**

Figure 5: Pressure predictions in normotension and disease cases (a) and (d) throughout the network at different time points. Yellow segments indicate regions where lesions are present.

**FIGURE 6**

Figure 6: Representative obstructed (red) and unobstructed (blue) pathways in the pulmonary vasculature, along with predictions of pressure and flow for the normotensive and four CTEPH disease scenarios. A zoom in of the flow predictions in the most distal vessels of the tree are shown to the right. Results are consistent with other pathways in the network.

*Small vessel hemodynamics.*

The structured tree model predicts pressure and flow in the smaller arteries and arterioles. Figure 7 shows normotensive predictions as well as CTEPH simulations in six large terminal arteries: three that are downstream to a lesion and three that are not. Mean pressure and flow are shown along the $\alpha$ and $\beta$ pathways, the longest and shortest pathways in the structured tree.

In the obstructed pathway, we see a significant drop in mean pressure and mean flow in case (a). This is consistent in both the $\alpha$ and $\beta$ pathways. In contrast, flow redistribution to the unobstructed vessels cause an increase in pressure and flow. Vasoconstriction of the terminal arteries and stiffening of the microvasculature in case (d) increases pressure yet flow in this pathway does not increase. Moreover, the mean pressure magnitudes in the unobstructed pathway are similar in both case (a) and case (d), though the systolic pressures (not shown) increase from 30 mmHg to 40 mmHg.

**FIGURE 7**

Figure 7: Hemodynamic predictions in the $\alpha$ and $\beta$ pathways of the structured tree in the normotensive case and CTEPH cases (a) and (d). (a) Large vessel network with terminal vessels in obstructed (T1-T3, red) and unobstructed (T4-T6, blue) pathways. (b) Schematic of the structured tree (grey branches) and the $\alpha$ and $\beta$ pathways (red and blue paths, respectively. (c) Mean pressure and flow predictions in the $\alpha$ and $\beta$ pathways in terminal vessels T1-T3 as a function of radius. (d) Mean pressure and flow predictions in the $\alpha$ and $\beta$ pathways in terminal vessels T4-T6 as a function of radius.

*Wave intensity analysis.*

Figure 8 presents wave intensity profiles in the MPA, LPA, and RPA for the normotensive and four disease cases. The wave reflection coefficient for each case is reported in Table 4. Normotensive simulations have a predominant FCW in all three branches, followed by a minor BCW. Overall, normotensive expansion waves are negligible. The wave reflection coefficient in the MPA is 0.47, smaller than the LPA and RPA (0.76 and 0.89). Adding lesions in case (a) leads to a prominent second FCW in the MPA and LPA, while RPA forward and backward compression waves have a bimodal profile that occur consecutively. Case (a) also produces a distinct bimodal BCW in both the MPA and LPA. In all three branches, both the forward and backward expansion waves are more oscillatory. The wave reflection coefficient increases in case (a) in the MPA and LPA but decreases in the RPA.

Small vessel stiffening in case (b) dampens expansion wave oscillations. The wave intensity profile is similar to case (a) but shows an elevated FCW across the three branches. A diminished BCW magnitude decreases the MPA and RPA reflection coefficients; the LPA reflection coefficient is the same in case (a) and (b). Downstream narrowing in case (c) has a negligible effect on the wave intensities, but slightly intensifies BCW magnitudes, increasing the reflection coefficient in all three vessels. For case (d), the most severe CTEPH, the peak wave intensity increases for all four wave types and the bimodal FCW and BEW are gone. Though the compression wave peak increases, the wave reflection coefficient decreases in all three vessels, suggesting a greater increase in magnitude for the FCWs than the BCWs.

**FIGURE 8**

Figure 8: Wave intensity analysis (WIA) at the midpoint of the main, left, and right pulmonary arteries (MPA, LPA, and RPA, respectively). Forward and backward running compression waves (FCW, BCW) are distinguished from forward and backward running expansion waves (FEW, BEW).

**Table 4**

Table 4: Wave reflection coefficient in the main, left, and right pulmonary arteries (MPA, LPA, and RPA) in each disease case. The coefficient, $\mathcal{R}$, is calculated using equation (27).

*BPA therapy.*

BPA therapy for inoperable CTEPH typically addresses 3-5 lesions per session; therefore, we target 3 lesions for the virtual BPA. This provides $\binom{20}{3} = 1140$ possible lesion combinations, which are all analyzed. We choose the optimal BPA using a combination of MPA pressure, WIA results and flow heterogeneity (KL-divergence) and want to minimize all three quantities. These three metrics are shown in Figure 9 for both the pre- (case (d)) and post-BPA simulations. Results show that the optimal treatment targets three ring-like lesions: two in the right lower lobe and one in the left upper lobe (here, anatomical left and right are opposite to the left and right sides of the figure). Lung tissue perfusion maps show reperfusion to both regions, and the flow PDF shifts leftward towards a more uniform network flow. The KL-divergence decreases from 3.05 to 2.95, signifying a smaller distance between the control and post-BPA flow PDF. Systolic MPA pressure decreases from 45 to 42 mmHg after BPA, and mean pressure decreases from 22 mmHg to 19 mmHg, below the diagnostic PH cutoff. Lastly, wave intensity results illustrate a decrease in BCWs and an increase in FEWs post-BPA. The wave reflection index is reduced in all three proximal arteries after BPA, changing from 0.49, 0.83, and 0.67 to 0.48, 0.76, and 0.61 in the MPA, LPA, and RPA, respectively. The pre-BPA FFR in these lesions is 0.44, 0.32, and 0.30, which improves to 0.88, 0.76, and 0.59, respectively, post-BPA.

**FIGURE 9**

Figure 9: Pre- and post-BPA metrics used to select the most effective treatment strategies. A total of 1140 interventions are considered. (a) Network of vessels with untreated lesions (cyan) and lesions treated in the best BPA simulation (red) along with pre and post-BPA perfusion maps. The region of the tissue most affected by BPA is identified within the black, dash-dot boxes. (b) Flow probability distribution functions (PDFs) for the normotensive (black), pre-BPA (red), and post-BPA (blue). (c) Pressure in the MPA pre-BPA (red) and post-BPA (black). (d) Wave intensities in the MPA, LPA, and RPA, pre-BPA (solid line) and post-BPA (dotted line).

## DISCUSSION

Computational hemodynamics is emerging as a useful clinical tool for cardiovascular disease. While several previous studies employ modeling to quantify treatment strategies for systemic artery disease (5, 71), less attention has been given to pulmonary vascular disease. In this study, we solve a patient-specific, 1D hemodynamics model in the large pulmonary arteries constructed from a CT image. The large vessels are coupled to a fractal-based small vessel model, enabling hemodynamic predictions at multiple scales. Our previous studies utilize a similar approach (8, 43, 50), yet this study is the first to utilize these methods with pressure loss models to generate perfusion maps and study CTEPH and BPA treatment strategies.

*Large artery hemodynamics.*

We consider five scenarios, ranging from normotensive to severe CTEPH. Figure 4 shows that lesions alone, case (a), elevate pressures above 20 mmHg and into the PH range. This agrees with results reported by Burrowes et al. (7), who use a steady flow, Poiseuille flow model to predict hemodynamics with and without pulmonary emboli. They argue that pulmonary vasoconstriction and stiffening are necessary to elevate mean arterial pressures above 25 mmHg. We observe similar trends in cases (b) and (c), which introduce narrowing and stiffening of the small vessels. Though cases (a)-(c) increase the mean MPA pressure, typical CTEPH patients have a systolic MPA pressure > 55 mmHg (59). We approach this in case (d), which also dilates and stiffens the large pulmonary arteries. In addition to elevated pressure, our results show that the LPA/RPA flow imbalance increases with disease severity. This parallels results by Spazzapan et al. (56), which compare simulations between stenosed and non-stenosed geometries.

WSS in the MPA and LPA decreases in case (a), while the RPA shear stress increases. An elevated pressure without a change in wall stiffness leads to a larger area deformation, decreasing the flow velocity and WSS in the MPA. Case (a) leads to a larger flow and WSS in the RPA but reduces the WSS in the LPA. Shear stress does not increase significantly in case (d), a result of the proximal artery dilation which increases area (in the denominator of equation (6). Results by Yang et al. (69) show a reduced time-averaged wall shear stress in pediatric patients with severe PH, congruent with our findings. Another study by Tang et al. (60) shows that proximal and distal WSS magnitude is smaller in PAH when compared to normotensive measurements, a consequence of MPA, LPA, and RPA dilation, also shown in this study. Though the mechanisms of PAH and CTEPH are different, both lead to proximal artery dilation (51, 69) and distal vascular narrowing (38, 51), hence a comparison of our results with PAH and CTEPH is appropriate.

A benefit of our model is the ability to project 1D network predictions onto surrounding lung tissue. Figure 3 shows a conversion of 1D model simulations to a 3D flow map, providing a meaningful and qualitative tool for physicians studying the effects of CTEPH. A recent study by Clark et al. (10) use a transmission line model to simulate large vessel hemodynamics and subsequently construct a lung perfusion map, similar to our work. However, their study did not account for perfusion deficits due to pulmonary lesions nor did it quantify flow heterogeneity, both of which are done here. Our model couples nonlinear, large arterial hemodynamics to both a linearized small vessel fluid dynamics model and a pressure loss term, capturing the complex relationship between pulmonary lesions and pulmonary hemodynamics. Figure 3 demonstrates the capability of our methodology in predicting large perfusion deficits in obstructed regions, as seen in vivo (22).

Our ring-like lesion model is based on the work by Young and Tsai (72) utilized in prior studies. The studies by Bezerra et al. (4) and Yin et al. (71) show that the combination of 1D CFD modeling with the aforementioned energy loss model is nearly identical to 3D stenosis model predictions in the coronary circulation. Only one prior pulmonary study (Spilker et al. (57)) utilized this energy loss model with a 1D framework in pigs; their results show that RPA stenosis leads to a drastic reduction of flow to the right lung and 25 mmHg increase in systolic pressure. However, none of these studies consider the effects of multiple distal lesions in an expansive pulmonary tree. In addition, we introduce a novel web-like lesion pressure loss term based on the Darcy-Forchheimer equation (70), which is a significant contribution to modeling CTEPH. These lesion types have not been modeled before, but are characteristic of a majority of CTEPH lesions (25).

Hemodynamics simulations throughout the pulmonary vasculature are feasible using our 1D CFD model framework. Figure 5 shows the mean pressure in each large pulmonary artery in both normotension and CTEPH, cases (a) and (d), during the cardiac cycle. While control simulations show a relatively consistent pressure throughout the vasculature at each time point, cases (a) and (d) illustrate a significant pressure drop downstream to lesions. This pressure drop, quantified by the FFR, is measured in CTEPH patients in two previous studies (22, 64) using optical coherence tomography. These studies report FFR values ranging from 0.22 to 0.90, similar to the FFR magnitudes provided here. Our results in Figure 2 show that ring like lesions have a larger pressure drop than web-like lesions, but that both lesion types have varied effects on MPA pressure. Pressure drops across different types of pulmonary lesions have only recently been documented (22, 25), and these studies suggest that web-like lesions are more common and easier to treat with BPA than other lesion types. These studies also show that the changes in pressure and flow downstream of web-like lesions are more volatile than other lesion types. To date, the properties of web-like lesions have not been investigated, but more information regarding the structure and composition of web-like lesions may be used to inform our loss model. Overall, our model shows that both lesion types contribute to a decrease in pressure downstream, with ring-like lesions playing a larger role.

Current hypotheses suggest that secondary arteriopathy is a determining factor in CTEPH recovery (14, 31, 38). As shown in Figure 6, model predictions in both obstructed and unobstructed pathways are affected in the different disease cases. Clearly, blood pressure drops downstream from the lesion while resistance to flow increases, leading to under perfusion in these vessels. Conversely, flow and pressure both increase in the unobstructed pathways. The study by Lang et al. (32) claims that decreased flow, and hence decreased shear stress, promotes pulmonary arterial remodeling downstream to a lesion. The authors also argue that increased flow and shear stress in unobstructed pathways leads to flow vasculopathy and, consequentially, pulmonary vascular remodeling. This remodeling of both obstructed and unobstructed arterial pathways explains why all small arterial vessels, i.e. with an area $\leq 5$ mm$^2$, narrow in CTEPH (31, 51). Our results agree with these hypotheses and illustrate under perfusion in obstructed pathways and hyper-perfusion in unobstructed pathways. Predictions in the obstructed path do not return to their original normotensive values, but they gradually approach these values as we introduce stiffening and narrowing in cases (b) through (d). Simulations from case (d) are consistent with the physiological conclusions that stiffening and narrowing of the pulmonary tree partially corrects pressure and flow imbalances due to CTEPH lesions (32).

*Small vessel hemodynamics.*

Our multiscale approach utilizes the structured tree model to simulate arteriolar hemodynamics distal to the large subsegmental arteries. In CTEPH, lesions in segmental and subsegmental arteries decrease flow to the microvasculature, leading to small vessel remodeling (23, 31). We simulate this in cases (b), (c), and (d) by decreasing the area of the terminal vessels and, consequently, the radii and number of branches in the structured tree. Results in Figure 6 show hemodynamic predictions in three structured trees in an obstructed path (T1, T2, and T3) and three structured trees in an unobstructed path (T4, T5, and T6). As expected, flow and pressure decrease substantially in the obstructed pathway with the addition of vascular lesions in case (a), whereas flow redistribution leads to an increase in flow and pressure in unobstructed terminal vessels and their arteriolar beds. Flow within the obstructed pathways does not increase when stiffness and PVR ($Z_{trm}$) are increased, though flow in the unobstructed pathway decreases as a result of stiffening. These results and methods agree with previous physiological studies. For example, the study by Stam et al. (58) induced CTEPH in swine, showing that animals with pulmonary lesions have increased wall-thickness and decreased microvascular luminal area compared to control. We model both these phenotypes of CTEPH in cases (b), (c), and (d), increasing MPA pressure. The review by Lang et al. (31) also supports the idea that an increase in flow in unobstructed pathways leads to secondary arteriopathy and wall thickening in the pulmonary arterioles.

Decreasing the radii of terminal arteries leads to a smaller arteriolar tree, shown in Figure 7 for case (d), as $r_{min}$ is held constant. A reduced microvascular density, also called microvascular rarefaction, is a known consequence of both CTEPH and PAH (9). An imaging study by Come et al. (51) concludes that CTEPH patients have a reduced volume distribution of small arteries when compared to control, and suggests pruning of the pulmonary arterioles. This phenomenon is modeled by Olufsen et al. (43) who simulate rarefaction of the pulmonary microvasculature and consequently predict increased mean and diastolic pressure in the MPA. The study by Yang et al. (69) uses a 3D-0D coupled model of the proximal and distal vasculature and predicts increased right ventricular pressure and distal WSS as a consequence of the pruned pulmonary microvasculature. However, these distal vascular predictions, as well as our own here, are hypothetical, as current imaging technologies cannot capture dynamics in vessels of this magnitude. This model type can test large and small vessel disease hypotheses and can be validated if measurements are available in both the large and small pulmonary arterial vessels.

*Wave intensity analysis.*

WIA is increasingly recognized as a measure of pulmonary vascular function (53, 59). Results in Figure 8 show how the addition of lesions and vascular stiffening affects forward and backward traveling waves. Results from the normotensive simulations show a large FCW in the MPA during systole followed by a smaller magnitude BCW and FEW, which rapidly decay in magnitude during diastole, and that the BEW is relatively negligible. In contrast, wave intensity magnitudes in scenario (d) nearly double in all four wave types. This agrees with the study by Su et al. (59), who measure area-velocity signals in both normotensive and PH patients, and show that wave intensity magnitudes nearly double in PH. Similarly, they show that FCWs in PH are approximately twice the magnitude of FEWs and BCWs, and that BEWs do not dissipate as quickly as in the normotensive patients. A similar conclusion is drawn in the study by Lau et al. (34); PAH patients in the study have a much larger BCW than control subjects, ultimately leading to an increased wave reflection coefficient. In this study, we define the wave reflection coefficient as the sum of forward and backward compression waves in the time domain. However, the wave reflection coefficient can be calculated using other methods (41, 59), including impedance analysis, either in the frequency or time domain (49).

Our results suggest that wave reflection coefficients increase with disease severity. However, dilating proximal vessels leads to a decreased reflection coefficient in the MPA, LPA, and RPA, as shown in Table 4. Though the reflection index is decreased in case (d), BCW magnitude increases significantly. This is shown to correlate with right ventricular dysfunction in the study by Schafer et al. (52). The authors of this study also argue that BCWs are more indicative of increased proximal stiffening, agreeing with our results in case (d) after stiffening the large proximal arteries. The alignment between our disease scenarios and previously published in-vivo WIA suggest that this model framework is suitable for addressing wave-propagation in PH and, more specifically, CTEPH. It is reported that proximal arteries both stiffen and dilate in long term PH (51), which is accounted for in case (d) in this study. Our simulations show that this physiological phenomenon decreases the wave reflection coefficient. This suggests that dilation of the MPA attempts to reduce the load on the heart by minimizing the mismatch between right ventricular ejection and reflected waves from the arterial periphery.

*Computational treatment planning.*

For inoperable CTEPH patients, a combination of drug therapy and BPA is the best alternative for improving vascular function. BPA is a physician dependent strategy and varies with both lesion location and severity of the disease (67). One common treatment strategy is to target the lobe with the largest perfusion deficit (37), yet there may be multiple lesions within each lobe. For this reason, our integrated mathematical model of CTEPH hemodynamics with patient-specific imaging can rank lesion importance in procedural planning. If integrated in the clinic, this model would allow cardiologists to solidify which lesions or lobes to address and could reduce uncertainty when deciding between 2-3 possible initial interventions. Image-based computational hemodynamics modeling is already recognized as a useful tool in surgical planning for coronary artery disease (5, 71), and this study is a first step in utilizing the same framework for understanding CTEPH and hemodynamic improvement after BPA intervention.

We utilize three indices to determine the best treatment strategy: the mean MPA pressure, the wave reflection coefficient in the MPA, and the KL-divergence of the flow field. The former two quantify how BPA affects vascular-ventricular coupling and ventricular afterload, while the latter quantifies improvements in perfusion. Our results show that a combination of all three indices leads to a best treatment for improving proximal hemodynamics and lung perfusion. Within our treatment framework, we reduce stenosis severity through $A_s$ for ring-like lesions or $K_{perm}$ for web-like lesions, but do not remove lesions. This is important to consider, as BPA does not remove the lesions completely (22, 31). In addition to targeting the lesions directly, we also decrease the degree of vasoconstriction in vessels downstream from a lesion, returning their radii to 90% of their original value. This is supported by the study by Boulate et al. (6) who show that distal artery thickness in pigs can regress to typical values seen in shams after removing lesions. Their study also analyzes arterial thickness 6 weeks after reperfusing the obstructed lung tissue, demonstrating that surgical removal of clots reestablishes a normal lumen area after surgery.

Results shown in Figure 9 illustrate that the combination of pressure predictions, WIA, and flow heterogeneity lead to an optimal BPA treatment. The optimal treatment here reperfuses the right lower lobe and restores some flow to the periphery in the right middle lobe and left upper lobe. Though physicians typically target the lobes of the lung with the largest perfusion deficit seen on V/Q scans (31), advances in new imaging technologies (such as optical coherence tomography) also assist in determining which lesions will best improve hemodynamics (25, 27). Our virtual BPA targets a region that is under perfused but does not identify the most flow deficient region (the left lower lobe) as the optimal location. There are several lesions located in this lobe, hence addressing only 3 lesions has little effect on redistributing the flow. Our optimal BPA treatment also identifies proximal lesions rather than distal lesions, consistent with clinical practice (31). Panel (b) in Figure 9 displays the relative cardiac output PDF for the normotensive control and pre- and post-BPA simulations, which are used to compute the KL-divergence. The control PDF has a bimodal structure, signifying two levels of perfused tissue as seen in Figure 3. Both pre- and post-BPA have a similar trend but show a small area of lung that is hyperperfused relative to normotensive simulations.

Our simulation results show that both MPA pressure and wave intensities change only minutely in response to the single BPA treatment. However, improvements in hemodynamics post-BPA are typically reported over the course of months or years and after multiple sessions (31, 36). Improvements immediately after BPA have only been reported in terms of FFR improvement (22, 37). Our optimal BPA treatment increased FFR from 0.44, 0.32, and 0.30 to 0.88, 0.76, and 0.59, respectively. These improvements are similar to recordings from the study by Ishiguro et al. (22), which discover an improved FFR after BPA. Their study shows that FFR increases from 0.22 to 0.59 and 0.34 to 0.86 in two different lesions, similar in magnitude to our results. The long-term effects of BPA likely include decreased vasoconstriction in small arteries and normalized wall thickness (6), but only in the case of reversible PH. Recent investigations into success rates in CTEPH treatment find that the severity of small vessel disease correlates negatively with successful surgical outcomes, suggesting that the degree of small vessel arteriopathy dictates whether PH is persistent after intervention (19). Histological data on arterioles after BPA would provide insight into the remodeling process after surgery and could be reflected in our simulations of post-BPA hemodynamics.

*Limitations and future work.*

We acknowledge several limitations in this study. We compute the wave reflection coefficient as the ratio of the cumulative forward and backward compression pressure waves. This is only one possible way of computing the wave-reflection coefficient in the time-domain (41), and can also be calculated in the frequency domain (49). It is unclear which index best captures features of the vasculature, yet all these metrics would support our conclusion that large vessel stiffening and the presence of pulmonary lesions increase BCWs and wave reflections.

This study implements a novel web-like lesion energy loss model; however, it is unclear if this model captures proximal and distal hemodynamic features in human web-like lesions. Future studies will be devoted to validating this model type by obtaining catheter measurements prior to and after web-like lesions (25). Moreover, we plan to investigate the sensitivity of model outputs, such as the wave reflection coefficient, mean MPA pressure, and flow heterogeneity index, to lesion and boundary condition parameters. In this manner, we will determine which parameters influence outputs after BPA and use this to guide parameter inference in patient-specific modeling studies for patients with CTEPH. We simulate both large arterial lesions and small vessel arteriopathy in this study, yet small vessel disease during the progression of CTEPH is not well understood. Several animal models have suggested mechanisms for pulmonary vascular remodeling (58), yet future animal studies may better illustrate the time-course of small vessel disease in CTEPH.

Though we provide hemodynamic predictions immediately after BPA and hypothesize some degree of vasodilation after an intervention, the short-term efficacy of a single BPA session is not yet clear. Data collection after single BPA procedures may provide more insight into the short-term remodeling process. To this point, we do not attempt to predict the long term remodeling of the pulmonary vasculature after intervention, which will be investigated in future studies via a growth and remodeling framework (62).

Flow mediated dilatation post-BPA likely contributes to improved hemodynamic function, and should be investigated further (32). This proof-of-concept study models CTEPH in a control geometry by placing lesions in the segmental and subsegmental branches. Future studies will integrate CTEPH images, including lesion location, type, and severity. Perfusion maps could also be compared to patient-specific V/Q scans, as well as quantitative data obtained from dual-energy CT (1). Other pulmonary vascular diseases, such as acute pulmonary embolism, PAH, or PH due to chronic pulmonary obstructive disorder (COPD), can also be modeled by integrating patient imaging data with our multiscale model. Lastly, we do not explicitly model the pulmonary capillaries or veins, which remodel with disease. Remodeling of the bronchiolar arteries is a common phenotype of CTEPH (32) and could be incorporated in future modeling studies.

## CONCLUSIONS

This study provides a framework that integrates patient-specific CT imaging with both large and small vessel fluid dynamics to predict multiscale hemodynamics in CTEPH. We model normotensive hemodynamics and four CTEPH disease cases, the latter driven by physiological hypotheses including large and small vessel remodeling. We utilize two lesion models representing ring-like and web-like lesions, common in CTEPH and imperative for simulation studies. Our results show that combining thromboembolic lesions with pulmonary vascular remodeling increases pulmonary pressures to the CTEPH range and mimics clinical observations. We predict perfusion in the lung tissue and provide a novel, quantitative metric for measuring perfusion heterogeneity, which is essential in understanding the link between flow deficits in the lung and disease severity. WIA results from the model framework are akin to prior measurements of wave intensities in CTEPH patients. Small vessel predictions are in agreement with clinical knowledge of CTEPH progression, and illustrate that lesions lead to downstream under perfusion as well as hyper perfusion in unobstructed territories. We propose a combination of indices predicted by the model and utilize these in prioritizing lesions for BPA therapy. Our modeling framework shows improvements in hemodynamics post-BPA, laying the foundation for future patient-specific investigations and validation studies using CTEPH data.

## APPENDIX

Writing pressure and flow as periodic, discrete, frequency domain functions, $P(x, \omega_k)$ and $Q(x, \omega_k)$, the linearized mass conservation and momentum balance equations for each frequency $\omega_k$ are given by

$$i\omega_k CP + \frac{\partial Q}{\partial x} = 0, \tag{29A}$$

and

$$i\omega_k Q + \frac{A_0\left(1 - F_J\right)}{\rho}\frac{\partial P}{\partial x} = 0. \tag{30A}$$

The quantity $F_J$ is the quotient of first and zeroth order Bessel functions

$$F_J = \frac{2J_1\left(\frac{rw_0}{r_0}\right)}{w_0 J_0(w_0)}, \quad w_0 = \sqrt{i^3 w^2}, \tag{31A}$$

where $w = \sqrt{r_0^2 \omega_k v}$ is the non-dimensional Womersley number. Equation (29A) depends on the vessel compliance $C$, approximated by

$$C = \frac{\partial p}{\partial A} \approx \frac{3}{2}\frac{r_0}{Eh}, \tag{32A}$$

where $Eh/r_0$ has the same form as equation (8), but with parameters $k_1^{sa}, k_2^{sa}$, and $k_3^{sa}$.

Differentiating equation (29A) with respect to $x$ and solving for $P$ in equation (30A) gives

$$\frac{\omega_k^2}{c}Q + \frac{\partial^2 Q}{\partial x^2} = 0, \quad c = \sqrt{\frac{A_0\left(1 - F_J\right)}{\rho C}} \tag{33A}$$

where $c$ (cm/s) is the pulse wave propagation velocity. Solving the above wave equation for $Q$ and plugging into equation (30A) gives

$$Q(x, \omega_k) = a\cos(\omega_k x/c) + b\sin(\omega_k x/c), \tag{34A}$$

$$P(x, \omega_k) = \frac{i}{g_\omega}\left(-a\sin(\omega_k x/c) + b\cos(\omega_k x/c)\right), \tag{35A}$$

where $a$ and $b$ are unknown integration constants and $g_\omega = cC = \sqrt{CA_0\left(1 - F_J\right)/\rho}$.

Using these definitions, vascular impedance is computed as

$$Z(x, \omega_k) = \frac{P(x, \omega_k)}{Q(x, \omega_k)} = \frac{i(-a\sin(\omega_k x/c) + b\cos(\omega_k x/c))}{g_\omega(a\cos(\omega_k x/c) + b\sin(\omega_k x/c))}. \tag{36A}$$

The above formulation is used to compute impedance at the inlet ($x = 0$) and outlet ($x = L$) of each small artery in the structured tree

$$Z(0, \omega_k) = \frac{i}{g_\omega}\frac{b}{a}, \tag{37A}$$

$$Z(L, \omega_k) = \frac{i(-a\sin(\omega_k L/c) + b\cos(\omega_k L/c))}{g_\omega(a\cos(\omega_k L/c) + b\sin(\omega_k L/c))}, \tag{38A}$$

Combining these equations gives the impedance at $x = 0$

$$Z(0, \omega_k) = \frac{i g_\omega^{-1} \sin(\omega_k L/c) + Z(L, \omega) \cos(\omega_k L/c)}{\cos(\omega_k L/c) + i g_\omega Z(L, \omega_k) \sin(\omega_k L/c)}. \tag{39A}$$

At the end of the structured tree, the distal impedance is set equal to the terminal impedance, $Z(L, 0) = Z_{trm}$. This impedance is then used to back solve the impedance values all the way through the root-impedance at the terminal end of the large arteries.

**GRANTS**

This study was funded in part by American Heart Association Grant #19PRE34380459 (Mitchel J. Colebank), American Heart Association Grant #19TPA34880033 (Sudarshan Rajagopal), NSF-DMS 1615820 (Mitchel J. Colebank and Mette S. Olufsen), and NIH-HL147590-01 (Mette S. Olufsen).

**ENDNOTE**

At the request of the authors, readers are herein alerted to the fact that additional materials related to this manuscript may be found at the website of the Cardiovascular Dynamics Group at North Carolina State University, which at the time of publication they indicate as: https://wp.math.ncsu.edu/cdg/. In particular, the 1D fluid dynamics model used in this study can be found on the Github website: https://github.com/mjcolebank/CDG_NCSU. These materials are not a part of this manuscript and have not undergone peer review by the American Physiological Society (APS). APS and the journal editors take no responsibility for these materials, for the website address, or for any links to or from it.

**Tables:**

**Table 1.** *Summary of disease scenarios simulated in this study.*

| Disease scenario | Description | Modeling approach |
|---|---|---|
| Normotensive | Typical pulmonary vasculature | - Low Stiffness<br>- Minimal PVR |
| (a) Lesions only | Mechanical obstructions due to ring- and web-like lesions | - 90% area reduction in 9 ring-like lesions<br>- 0.009 permeability in 11 web-like lesions |
| (b) Lesions with local remodeling | Increased vascular stiffness, narrowing of vessels distal to lesions, and increased PVR | - Increased stiffness in vessels distal to lesions<br>- 40% reduction in radius in vessels distal to lesions<br>- Increase PVR in terminal vessels distal to lesions |
| (c) Lesions with global small vessel arteriopathy | Narrowing, stiffening, and increased resistance in all small vessels | - 40% reduction in radius in vessels with $A_0 < 5\ \text{mm}^2$<br>- Increased stiffness in narrowed vessels<br>- Increased PVR in affected terminal vessels |
| (d) Lesions with proximal remodeling and small vessel arteriopathy | Narrowing and stiffening of small vessels, dilation and stiffening of large proximal vessels, and increased PVR | - Small vessel remodeling from (c)<br>- Dilation of proximal arteries with $A_0 \geq 10\ \text{mm}^2$<br>- Increased network stiffness<br>- Increased PVR in all terminal vessels |

**Table 2.** *Parameter values for simulating the five different disease scenarios described in the section Chronic Thromboembolic Pulmonary Hypertension (CTEPH). $k_3$ and $Z_{trm}$ values are listed for vessels in both the unobstructed and obstructed pathways (in parentheses).*

| | **Disease case** | | | | |
|---|---|---|---|---|---|
| Parameter | Normotensive | a | b | c | d |
| $k_1^{la}$ (g/ (cm s$^2$)) | $2.5 \times 10^6$ | $2.5 \times 10^6$ | $2.5 \times 10^6$ | $2.5 \times 10^6$ | $1.0 \times 10^7$ |
| $k_2^{la}$ (1/cm) | $-15$ | $-15$ | $-20$ | $-20$ | $-10$ |
| $k_3^{la}$ (g/ (cm s$^2$)) | $6.4 \times 10^4$ | $6.4 \times 10^4$ | $8.0 \times 10^4$ | $8.0 \times 10^4$ | $2.0 \times 10^5$ |
| | - | - | $(1.6 \times 10^5)$ | $(1.6 \times 10^5)$ | $(4.0 \times 10^5)$ |
| $k_1^{sa}$ (g/ (cm s$^2$)) | $2.5 \times 10^7$ | $2.5 \times 10^7$ | $5.0 \times 10^7$ | $5.0 \times 10^7$ | $5.0 \times 10^7$ |
| $k_2^{sa}$ (1/cm) | $-15$ | $-15$ | $-20$ | $-20$ | $-20$ |
| $k_3^{sa}$ (g/ (cm s$^2$)) | $8.0 \times 10^5$ | $8.0 \times 10^5$ | $1.6 \times 10^6$ | $1.6 \times 10^6$ | $1.6 \times 10^6$ |
| | - | - | $(1.6 \times 10^7)$ | $(1.6 \times 10^7)$ | $(1.6 \times 10^7)$ |
| $k^{lesion}$ (g/ (cm s$^2$)) | - | $6.4 \times 10^5$ | $8.0 \times 10^5$ | $8.0 \times 10^5$ | $2.0 \times 10^6$ |
| $Z_{trm}$ (g/ (cm$^4$ s)) | $1.0 \times 10^5$ | $1.0 \times 10^5$ | $1.0 \times 10^6$ | $1.0 \times 10^7$ | $1.0 \times 10^8$ |
| | - | - | $(1.0 \times 10^8)$ | $(1.0 \times 10^8)$ | $(1.0 \times 10^8)$ |

**Table 3.** *Parameters describing ring- and web-like lesions and their values.*

| Stenosis parameters | Description | Value |
|---|---|---|
| $K_t$ (ND) | Turbulent pressure loss coefficient | 1.52 |
| $K_u$ (ND) | Inertial pressure loss coefficient | 1.2 |
| $L_s$ (cm) | Ring-like lesion length | $L_i/4$ |
| $K_{perm}$ (cm$^2$) | Darcy loss coefficient for web-like lesion permeability | 0.009 |
| $G$ (cm) | Inertial loss coefficient for web-like lesions | 0.10 |
| $L_w$ (cm) | Length of web-like lesion | $L_i/2$ |

**Table 4.** *Wave reflection coefficient in the main, left, and right pulmonary arteries (MPA, LPA, and RPA) in each disease case. The coefficient, $\mathcal{R}$, is calculated using equation (27).*

| Wave reflection coefficient ($\mathcal{R}$) | | | |
|---|---|---|---|
| Disease case | MPA | LPA | RPA |
| normotensive | 0.47 | 0.76 | 0.89 |
| case (a) | 0.74 | 0.88 | 0.86 |
| case (b) | 0.68 | 0.88 | 0.83 |
| case (c) | 0.74 | 0.90 | 0.86 |
| case (d) | 0.49 | 0.83 | 0.67 |

**Figures:**

**Figure 1.** Schematic of model development, starting from image analysis and image segmentation and ending in construction of the skeletonized network and computational domain.

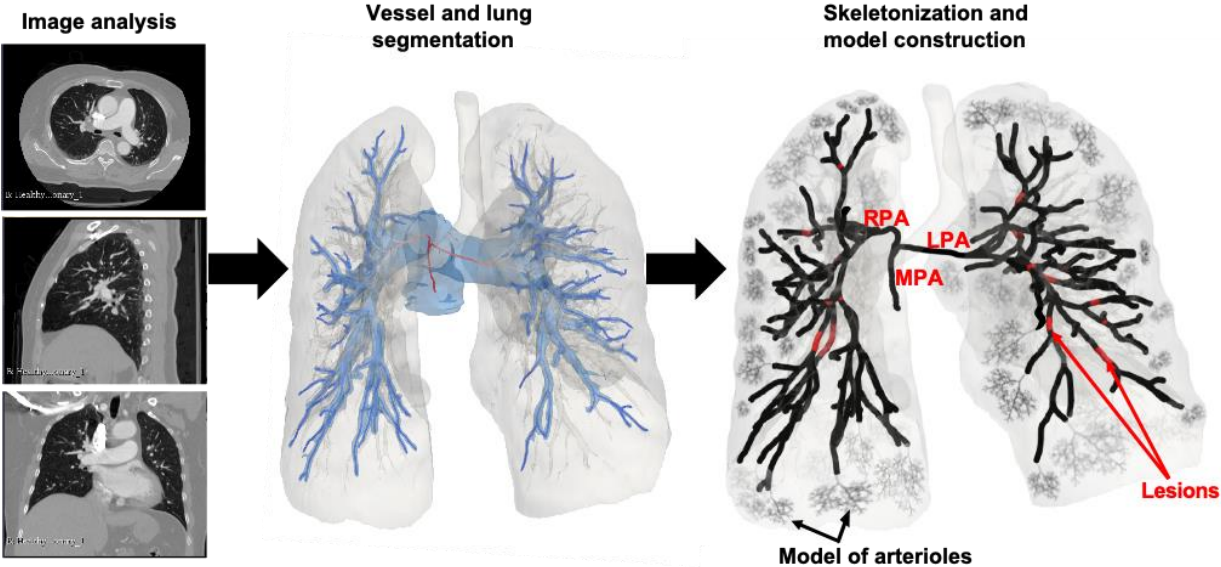

**Figure 2.** Predictions of mean MPA pressure (top), and fractional flow reserve (FFR, bottom) when increasing lesion severity in ring-like (left) or web-like (right) lesions. FFR predictions correspond to 9 ring-like lesions and 11 web-like lesions. Overall, ring-like lesions have a greater effect on MPA pressure and FFR than web-like lesions.

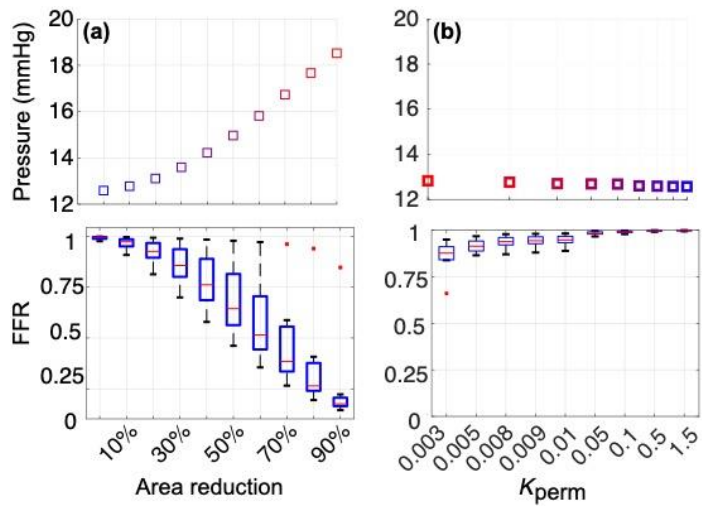

**Figure 3.** Perfusion maps for increasing ring-like stenosis severity. The top left panel (0%) shows the normotensive perfusion map, while the bottom right perfusion map shows 90% severity. The Kullback–Leibler (KL) divergence, constructed from the flow probability density function (PDF), quantifies the flow heterogeneity. PDFs are constructed using kernel density estimation with a bandwidth of 0.009 and a reflection boundary correction.

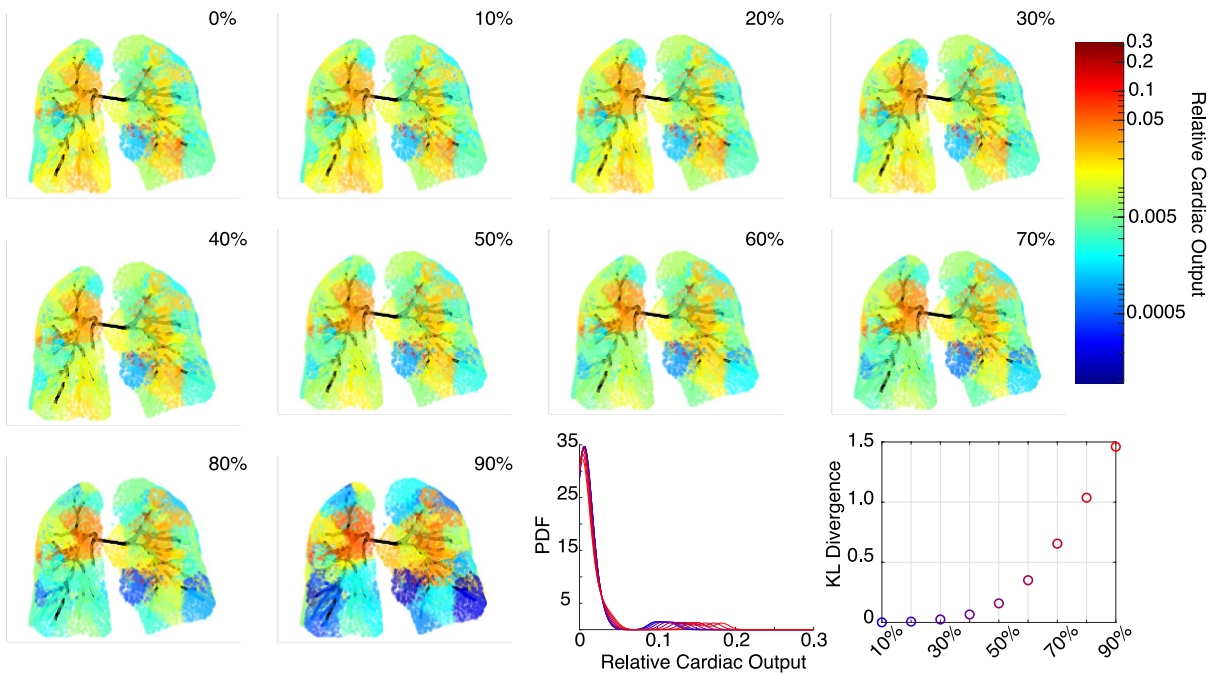

**Figure 4.** Pressure, flow, and shear stress in the main, left, and right pulmonary arteries in each disease scenario. Simulations in case (a) include vascular lesions without large or small vessel remodeling. Simulations in cases (b), (c), and (d) include additional local small vessel remodeling, global small vessel remodeling, and large vessel dilatation and stiffening, respectively.

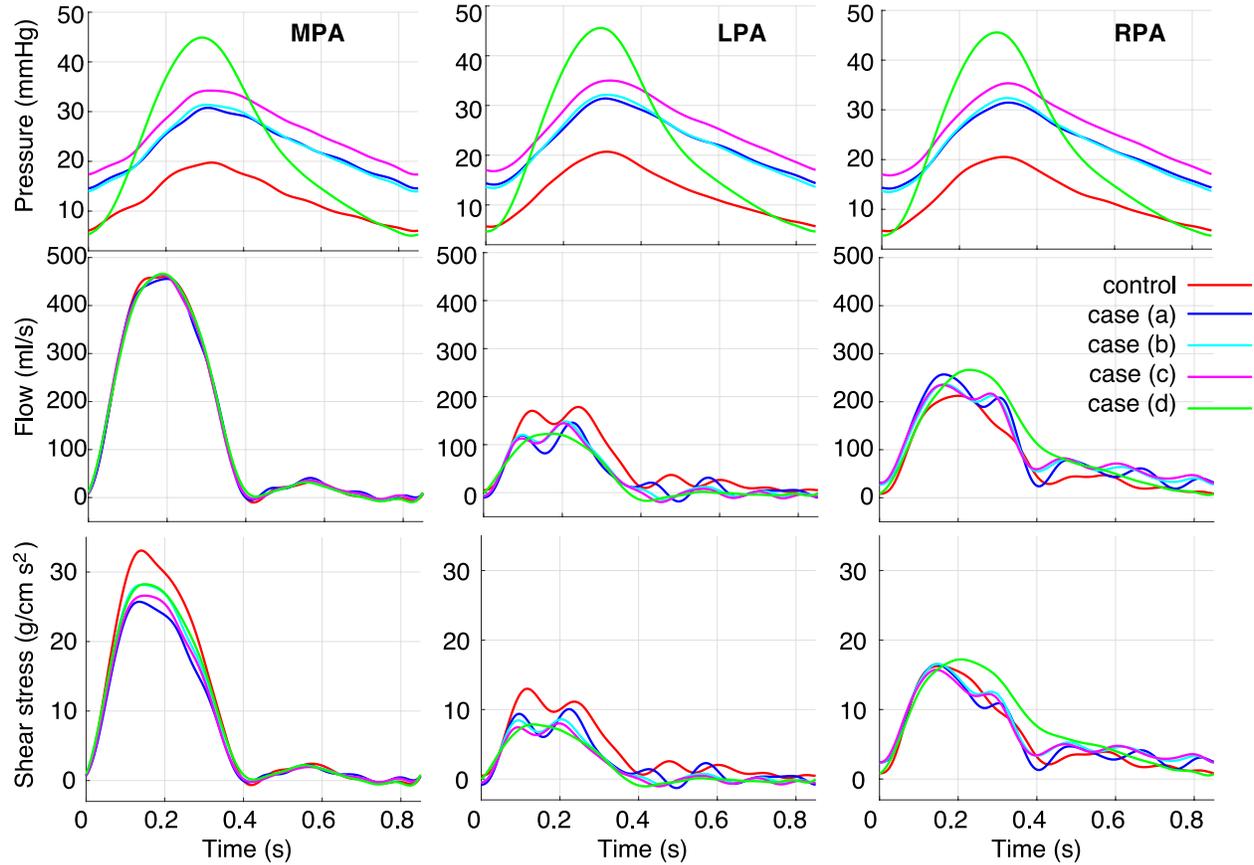

**Figure 5.** Pressure predictions in normotension and disease cases (a) and (d) throughout the network at different time points. Yellow segments indicate regions where lesions are present.

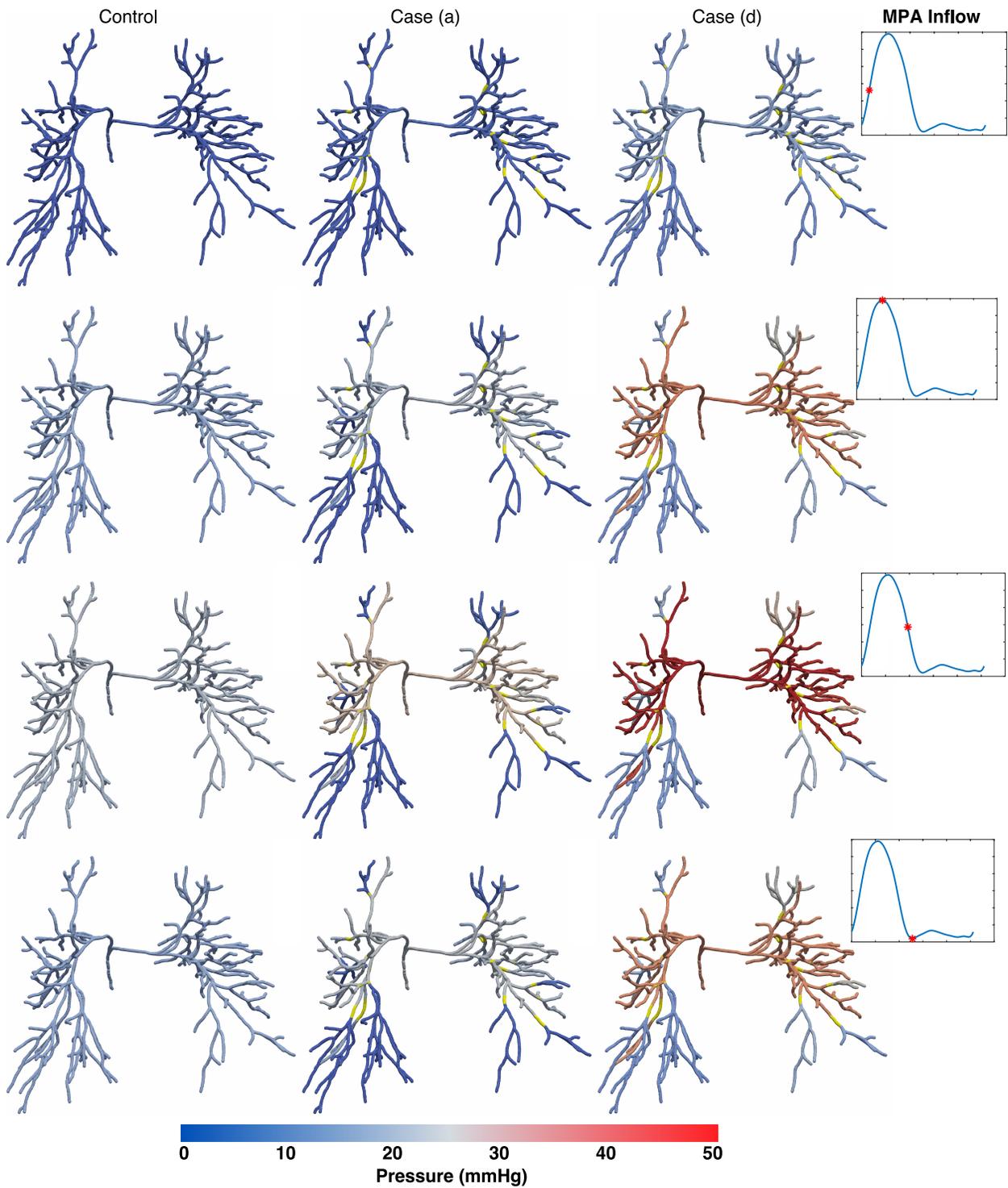

**Figure 6.** Representative obstructed (red) and unobstructed (blue) pathways in the pulmonary vasculature, along with predictions of pressure and flow for the normotensive and four CTEPH disease scenarios. A zoom in of the flow predictions in the most distal vessels of the tree are shown to the right. Results are consistent with other pathways in the network.

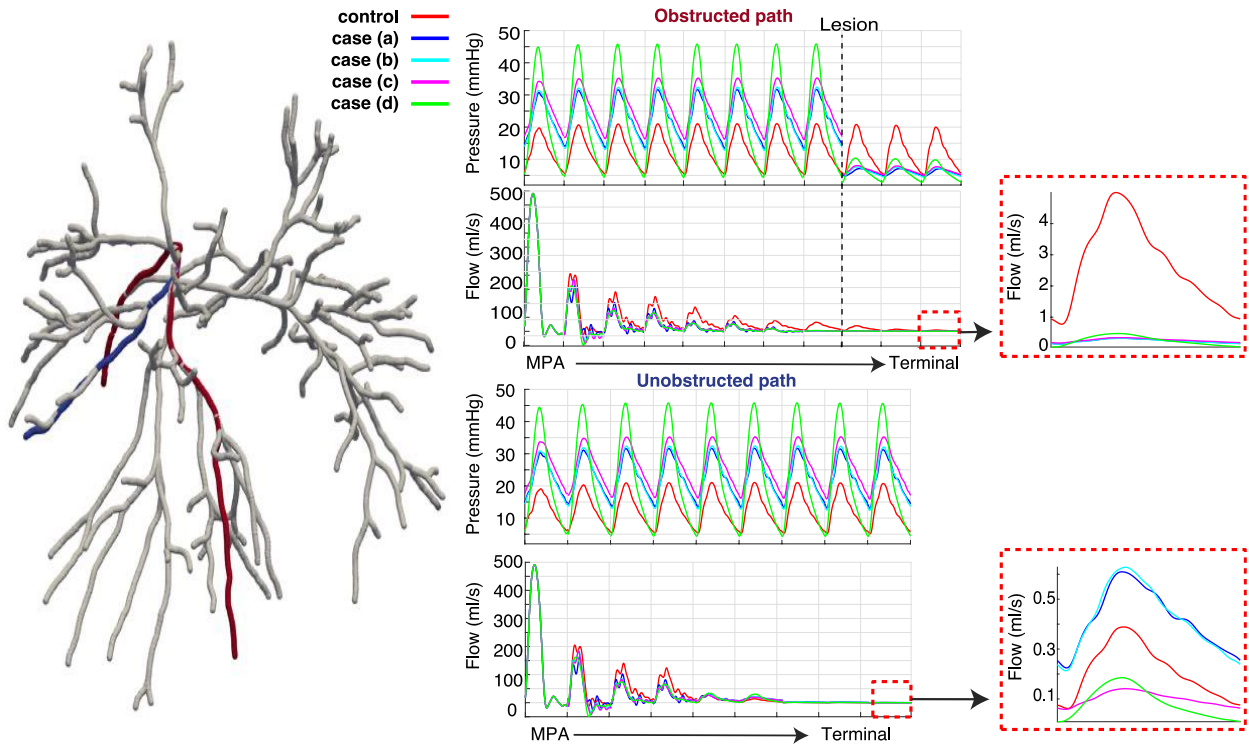

**Figure 7**. Hemodynamic predictions in the $\alpha$ and $\beta$ pathways of the structured tree in the normotensive case and CTEPH cases (a) and (d). (a) Large vessel network with terminal vessels in obstructed (T1-T3, red) and unobstructed (T4-T6, blue) pathways. (b) Schematic of the structured tree (grey branches) and the $\alpha$ and $\beta$ pathways (red and blue paths, respectively. (c) Mean pressure and flow predictions in the $\alpha$ and $\beta$ pathways in terminal vessels T1-T3 as a function of radius. (d) Mean pressure and flow predictions in the $\alpha$ and $\beta$ pathways in terminal vessels T4-T6 as a function of radius.

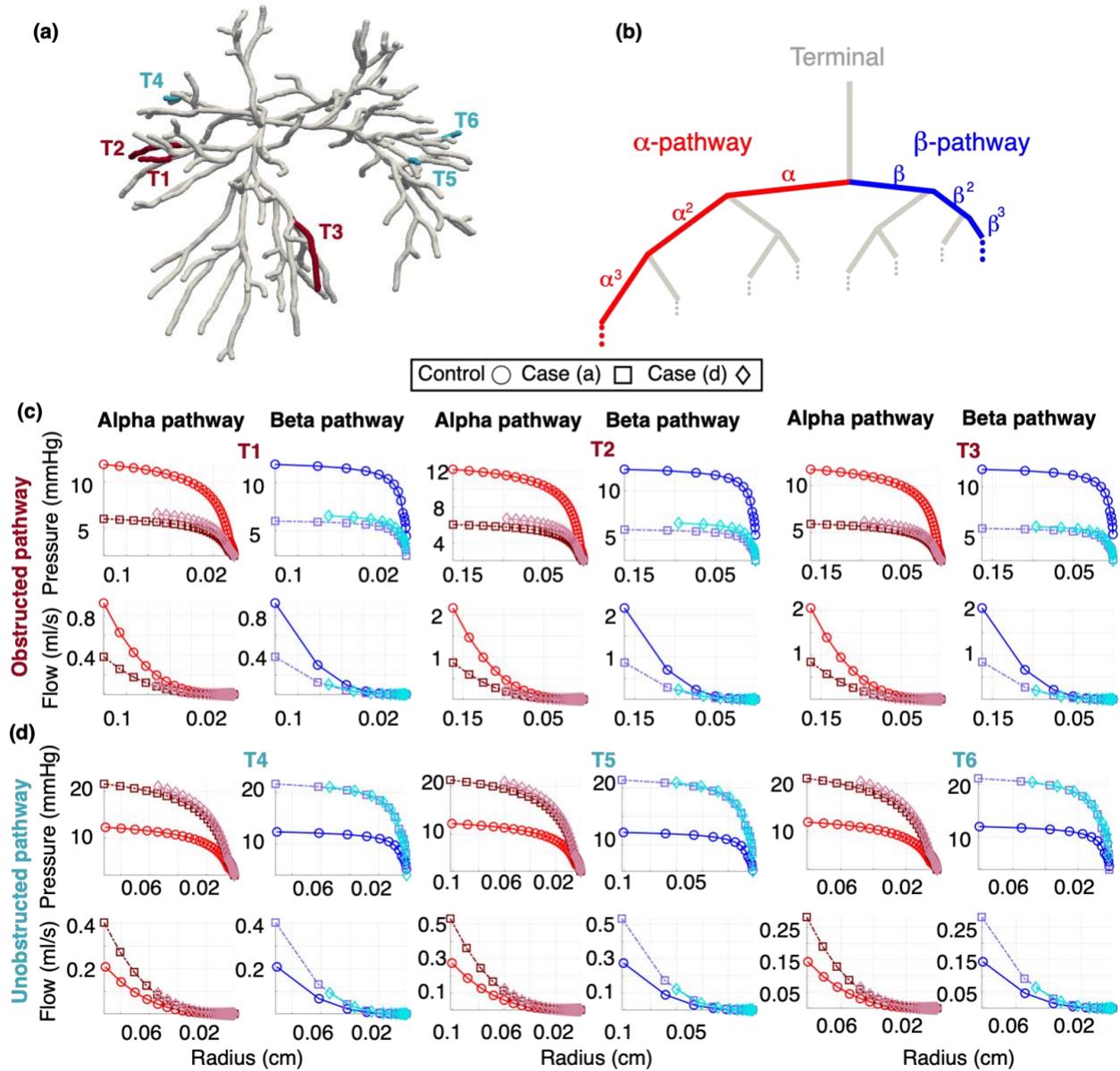

**Figure 8.** Wave intensity analysis (WIA) at the midpoint of the main, left, and right pulmonary arteries (MPA, LPA, and RPA, respectively). Forward and backward running compression waves (FCW, BCW) are distinguished from forward and backward running expansion waves (FEW, BEW).

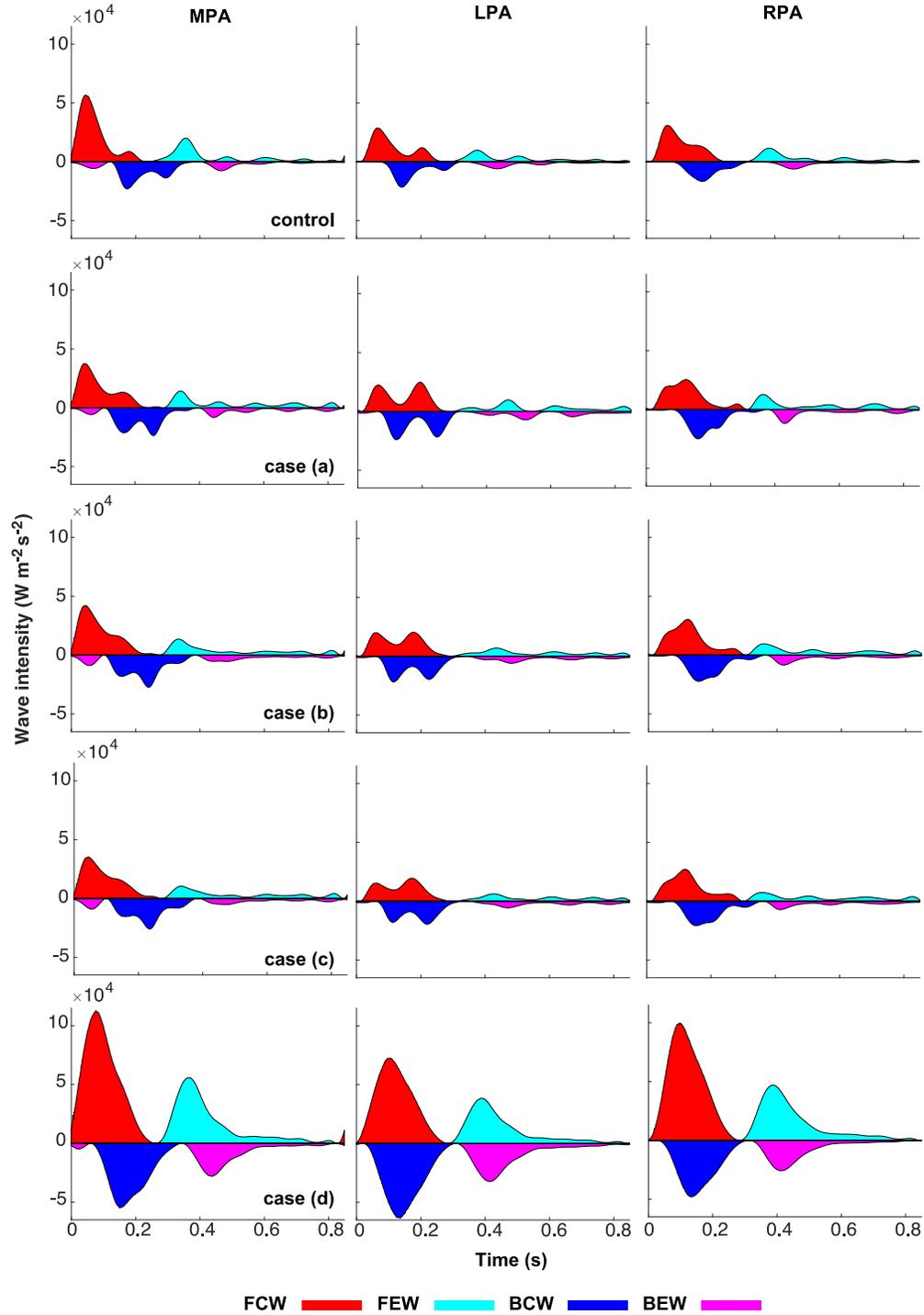

**Figure 9.** Pre- and post-BPA metrics used to select the most effective treatment strategies. A total of 1140 interventions are considered. (a) Network of vessels with untreated lesions (cyan) and lesions treated in the best BPA simulation (red) along with pre- and post-BPA perfusion maps. The region of the tissue most affected by BPA is identified within the black, dash-dot boxes. (b) Flow probability distribution functions (PDFs) for the normotensive (black), pre-BPA (red), and post-BPA (blue). (c) Pressure in the MPA pre-BPA (red) and post-BPA (black). (d) Wave intensities in the MPA, LPA, and RPA, pre-BPA (solid line) and post-BPA (dotted line).

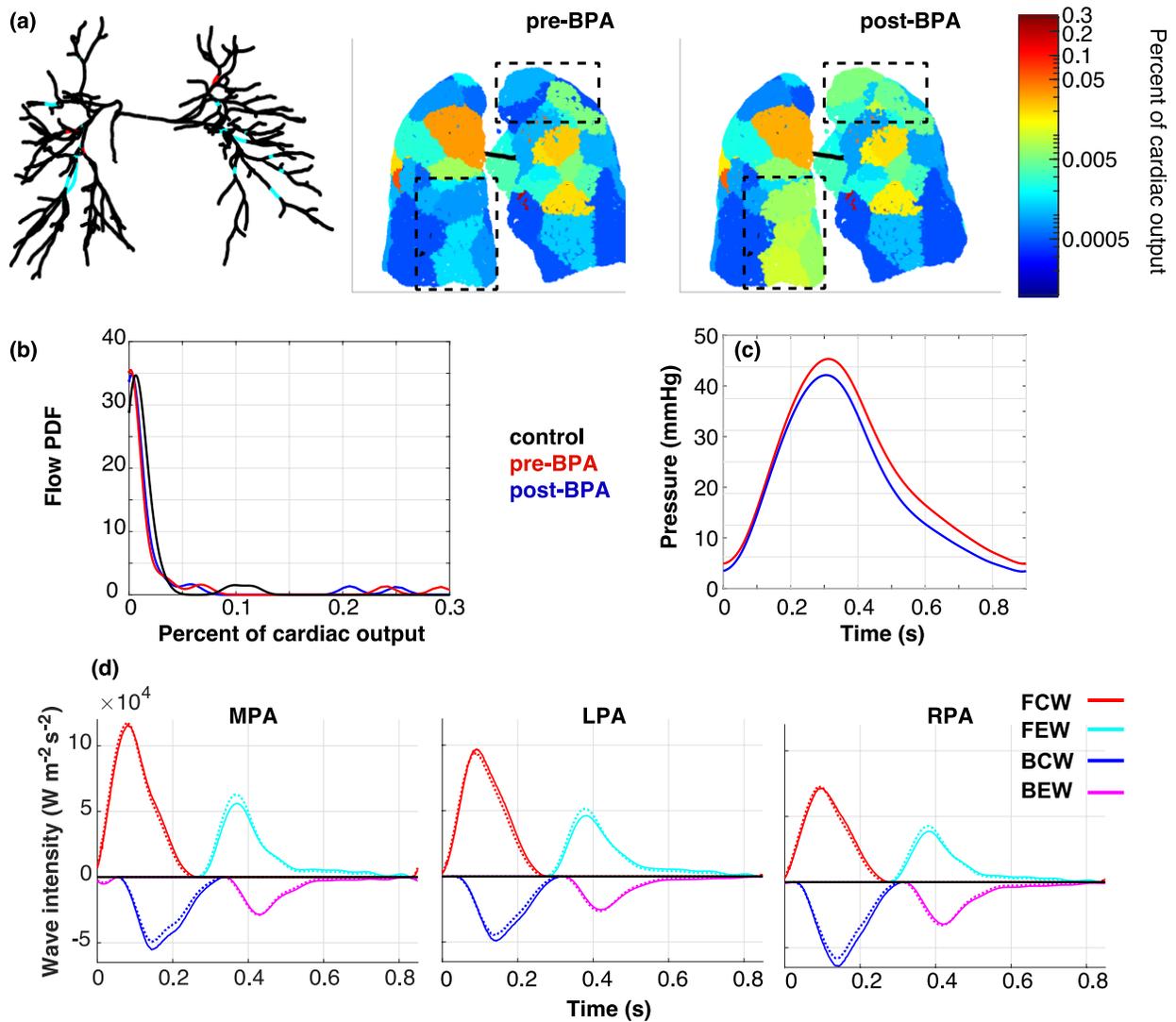